\documentclass[aps,showpacs,nofootinbib,preprintnumbers,amsmath,amssymb,twocolumn,superscriptaddress,prx,floatfix,nofootinbib]{revtex4-2}

\usepackage{amsfonts}
\usepackage[hidelinks]{hyperref}
\usepackage{xcolor}
\hypersetup{
    colorlinks,
    linkcolor={red!50!black},
    citecolor={blue!50!black},
    urlcolor={blue!80!black}
}
\usepackage{graphicx}
\usepackage{dcolumn}
\usepackage{bm}
\usepackage{xkcdcolors}
\usepackage{tabularx}
\usepackage{epstopdf}
\usepackage{xcolor}
\usepackage{mathrsfs}
\usepackage{subfig}
\usepackage{mathtools}
\usepackage{float}
\usepackage{algpseudocode}
\usepackage{verbatim}
\usepackage{comment}
\usepackage[capitalise]{cleveref}
\usepackage{ragged2e}

\DeclareCaptionJustification{justified}{\justifying}
\newcolumntype{C}[1]{>{\centering\arraybackslash}p{#1}}

\begin{document}

\title{Gravity Models of Networks:\\Integrating Maximum-Entropy and Econometric Approaches}

\author{Marzio Di Vece}
\email{marzio.divece@imtlucca.it}
\affiliation{IMT School for Advanced Studies Lucca, P.zza San Francesco 19, 55100 Lucca (Italy)}

\author{Diego Garlaschelli}
\affiliation{IMT School for Advanced Studies Lucca, P.zza San Francesco 19, 55100 Lucca (Italy)}
\affiliation{Lorentz Institute for Theoretical Physics, Leiden University, Niels Bohrweg 2, 2333 CA Leiden (The Netherlands)}
\author{Tiziano Squartini}
\affiliation{IMT School for Advanced Studies Lucca, P.zza San Francesco 19, 55100 Lucca (Italy)}
\affiliation{Institute for Advanced Study (IAS), University of Amsterdam, Oude Turfmarkt 145, 1012 GC Amsterdam (The Netherlands)}

\date{16th July 2021}

\begin{abstract}
The World Trade Web (WTW) is the network of international trade relationships among world countries. Characterizing both the local link weights (observed trade volumes) and the global network structure (large-scale topology) of the WTW via a single model is still an open issue.
While the traditional Gravity Model (GM) successfully replicates the observed trade volumes by employing macroeconomic properties such as GDP and geographic distance, it unfortunately predicts a fully connected network, returning a completely unrealistic topology of the WTW. To overcome this problem, two different classes of models have been introduced in econometrics and statistical physics. Econometric approaches interpret the traditional GM as the expected value of a probability distribution that can be chosen largely arbitrarily and tested against alternative distributions. Statistical physics approaches construct maximum-entropy probability distributions of (weighted) graphs from a chosen set of measurable, structural constraints and test distributions resulting from different constraints. Here we compare and integrate the two approaches by considering a class of maximum-entropy models that can incorporate macroeconomic properties used in standard econometric models. We find that the integrated approach achieves an overall better performance than the purely econometric one. These results suggest that the maximum-entropy construction can serve as a viable econometric framework wherein extensive and intensive margins can be separately controlled for, by combining topological constraints and dyadic macroeconomic variables.
\end{abstract}


\maketitle

\section{Introduction\label{secI}}

The World Trade Web (WTW), formed by the import/export relationships between world countries, has received a lot of attention both in traditional economic studies and in more recent investigations. Indeed, as recent crises have clearly pointed out, `network effects matter'~\cite{Schweitzer2009}: trade linkages are, in fact, the most important channels of interaction between world countries, directly transmitting economic shocks~\cite{Fagiolo2010b,Saracco2016,Squartini2015a} or indirectly transmitting financial ones~\cite{Schiavo2010,Squartini2013,Kali2007,Kali2010,Starnini2019}). It therefore comes with no surprise that many contributions have focused on the analysis of the \textit{network properties} of the WTW. Indeed, the (ever-increasing) data availability over the last years has motivated researchers (coming from disciplines as different as economics, network science and social sciences) to explore the architecture of the WTW from an empirical point of view, complementing the traditional theoretical knowledge in classical trade economics.

These efforts have produced a large wealth of literature characterizing various stylized facts of the WTW architecture~\cite{Barigozzi2010,Fronczak2012a,Fronczak2012b,Serrano2003,Garlaschelli2004,Garlaschelli2005,Fagiolo2010,Fagiolo2008a,Fagiolo2008b, Schweitzer2009, Vitali2011,Schiavo2010}.
A first stream of contributions has focused on the purely \emph{binary} properties of the WTW, i.e. the structural features that require the knowledge of only the presence of connections (trade relationships) between countries, irrespective of their intensity. These studies have highlighted a large (compared with most of the other real-world networks) density, a right-skewed and heavy-tailed degree distribution, disassortative mixing by degree (indicating that countries having many trade partners are, on average, connected with countries having few partners), a hierarchical organization (indicating that partners of well-connected countries are less interconnected than those of poorly-connected ones), the presence of a core of countries trading with almost everyone else, and finally the presence of (a sort of) bow-tie structure~\cite{Garlaschelli2004,Garlaschelli2005,Fagiolo2008b,Vitali2011,Jeroen2019}. 
A second stream of contributions has focused on the weighted properties of the WTW, where the \emph{weight} attached to each link represents the volume of the corresponding trade relationship and the \emph{strength} (sum of all link weights) of each node represents the overall trade volume of the corresponding country. These studies have highlighted right-skewed and heavy-tailed (generally log-normal) weight and strength distributions (indicating that few intense trade connections coexist with a majority of low-intensity ones), disassortative mixing by strength (indicating that countries whose trade volume is large are, on average, connected with countries whose trade volume is small), and a large weighted clustering coefficient (indicating that the trade volume of the partners of well-connected countries is larger than the trade volume of the partners of poorly connected ones)~\cite{Fagiolo2008a}.\\

From the modelling side, two broad classes of models of the international trade system can be roughly identified: \textit{econometric} models and \textit{network} models (the latter mainly rooted into statistical physics). Both can be traced back to the earliest model of international trade, proposed in 1962 by the physics-educated Dutch economist Jan Tinbergen (considered the founding father of econometrics together with Ragnar Frisch). Tinbergen modelled the import/export flows between countries via a relationship that is formally analogous to the law of gravity, where the `masses' of countries are replaced by GDPs and the inter-country distances are replaced by suitably defined geographic distances~\cite{Tinbergen1962}. This is the celebrated Gravity Model (GM) and can be shown to reproduce the positive trade volumes quite accurately~\cite{Anderson2011, Bergeijk2010,DeBenedictis2011,Almog2019}. 
Although our focus here is on international trade, it is worth mentioning that the GM has been used successfully to describe the positive link weights associated to processes relevant to many other networks as well, including migration flows~\cite{ravenstein,migrationGM} (which actually represent the earliest~\cite{ravenstein} application of the GM),
mobility and traffic patterns~\cite{zipf,balcan,koreaGM}, communication streams~\cite{LambiotteGM}, and spreading phenomena~\cite{Ma2016,Li2019}.
Despite its success, the most evident limitation of the GM consists in predicting that each country establishes a trading relationship with every other country, a result that is in stark contrast with empirical data, where `missing' trade relationships are actually found to be a significant proportion (up to one half) of the number of all possible pairs of countries. Since systematically overestimating the number of connections is known to lead to a significant miscalculation of network effects~\cite{Schweitzer2009}, one should avoid the use of the pure GM as a reliable network model of the WTW~\cite{Almog2019}.

From the Sixties on, a lot of work has been done in econometrics to overcome the aforementioned limitation. Initially, Eaton and Tamura~\cite{Eaton1995} and Martin and Pham~\cite{Martin2008} suggested to employ a tobit-like estimation procedure, by rounding to zero the trade flows below a certain threshold.
Later, the serious conceptual problem posed by the arbitrariness of the chosen threshold motivated Helpman, Melitz and Rubinstein~\cite{Helpman2008} to propose a two-step estimation procedure: first, a probit model is employed to estimate the probability of observing a trade relationship between any two nodes; subsequently, the corresponding trade flow is estimated via an Ordinary-Least-Squares (OLS) regression whose parameters are tuned on the entire set of positive weights.
An alternative algorithm is the one proposed by Silva and Tenreyro~\cite{Silva2006}, who estimated the gravity equation in a multiplicative - rather than additive - fashion by employing a Poisson Pseudo-Maximum Likelihood (PPML) method and obtained good estimates of trade flows, robust to heteroskedasticity. In the following years, PPML has been proven to be very sensitive to the `excess zeros' of the dependent variables. Hence, a different class of two-step estimation procedures has been devised, i.e. the so-called `zero-inflated' (ZI) methods~\cite{Burger2009,Winkelmann2008}: these models are defined by a logit estimation, aimed at establishing the presence of a link, followed by either a Poisson (ZIP) or a negative binomial (ZINB) regression, aimed at estimating the corresponding trade volume.
Later in 2011, Due\~nas and Fagiolo~\cite{Duenas2011} proved that the weighted structure of the WTW is well predicted by GMs if and only if its topological structure is specified as well. This result marks an important shift from a perspective whose only focus use to be the prediction of trade flows to a perspective where the topological structure of the network becomes one of the reconstruction targets.\\

Coming to the physics-inspired approaches to international trade, the majority of proposed models are maximum-entropy models of networks~\cite{Squartini2011a,Squartini2011b,Squartini2011c,Mastrandrea2014,Bargigli2014,DSBook,Almog2019}. The maximum-entropy (ME) framework~\cite{Jaynes1957a,Jaynes1957b,Jaynes1982,Cimini2019,Parisi2020,DSBook,Squartini2011b} allows for purely structural network properties to be constrained in order to derive the maximally unbiased probability distribution compatible with those properties. A series of publications over the years~\cite{Garlaschelli2004,Squartini2011a, Squartini2017,Squartini2015b,Cimini2021,Squartini2011c,Almog2019} has showed that fixing the degree sequence of world countries is enough to successfully reproduce higher-order binary properties such as the average nearest-neighbour degree and the clustering coefficient. By contrast, this result is no longer retrieved when the analysis is reformulated within a purely weighted framework: for instance, it turns out that the strength sequence is not an effective constraint in reproducing the average nearest-neighbour strength and the weighted clustering coefficient~\cite{Squartini2011b,Fagiolo2012}. In fact, accuracy in the reconstruction of weighted properties is recovered only if the degrees and the strengths are constrained jointly~\cite{Mastrandrea2014b,Garlaschelli2009,Gabrielli2019}.\\

Econometric and network approaches have mostly proceeded along parallel tracks, with little interaction so far. The only line of research that has systematically looked for the possibility of an integration between the two approaches is the one reformulating maximum-entropy network ensembles as \textit{hidden variable} (or \textit{fitness}~\cite{Caldarelli2002}) models~\cite{Garlaschelli2004,Garlaschelli2005,Almog2015,Almog2017,Almog2019}. In this approach, the maximum-entropy construction of the network ensemble is formally preserved, but the Lagrange multipliers usually viewed as free parameters tuned in order to enforce the chosen constraints are instead identified with empirical macroeconomic factors, most notably the GDP of countries. Building on those results, as well as on the success of physics-inspired models in reconstructing various kinds of socio-economic and financial systems~\cite{Cimini2021,Squartini2018}, here we explore the potential of the maximum-entropy formalism to provide a viable framework for econometrics too.\\

The rest of the paper is organized as follows. Section~\ref{secII} reviews the traditional GM and its performance in reproducing the positive WTW weights. Section~\ref{secIII} is devoted to the description of purely econometric models and their application to the analysis of trade (networks). Section~\ref{secIV} is devoted to the description of maximum-entropy network models and their application to the WTW. Section~\ref{secV} illustrates the main results of the paper, i.e. the integration of purely econometric and physics-inspired models. Section~\ref{secVI} concludes and presents an outlook on possible future extensions.

\section{Modelling positive weights\label{secII}}

From a merely econometric point of view, the simplest exercise is that of reproducing the realized (positive) trade volumes (link weights) of the WTW. To this aim, we have considered two different datasets. The first one is curated by Gleditsch~\cite{Gleditsch2002} and includes yearly trade volumes, yearly GDP values (both reported in millions of US dollars) and the (time-independent) matrix of geographic distances between capital cities of all countries in the data. The second one is the BACI dataset, a detailed description of which can be found in~\cite{baci1,baci2}. For both datasets we have selected and analyzed eleven years: 1990-2000 for the Gleditsch dataset and 2007-2017 for the BACI one. The year 2000 of the Gleditsch dataset - i.e. the one with the largest number of countries ($176$) - is the snapshot we have selected to graphically illustrate all the results of our analyses. We have always considered the undirected (symmetrized) version of the weighted trade matrix, whose generic entry reads $w_{ij}=\frac{\text{exp}_{ij}+\text{exp}_{ji}}{2}$, i.e. $w_{ij}$ is the bilateral trade volume defined as the arithmetic mean of the export volume from country $i$ to country $j$ and of the export volume from country $j$ to country $i$.

\begin{figure*}[t!]
\subfloat[\centering$\omega_i\omega_j$ ($\textcolor{xkcdPink}{\bullet}$) VS $\rho(\omega_i\omega_j)$ ($\textcolor{xkcdPurple}{\bullet}$)]
{\includegraphics[width=.31\textwidth]{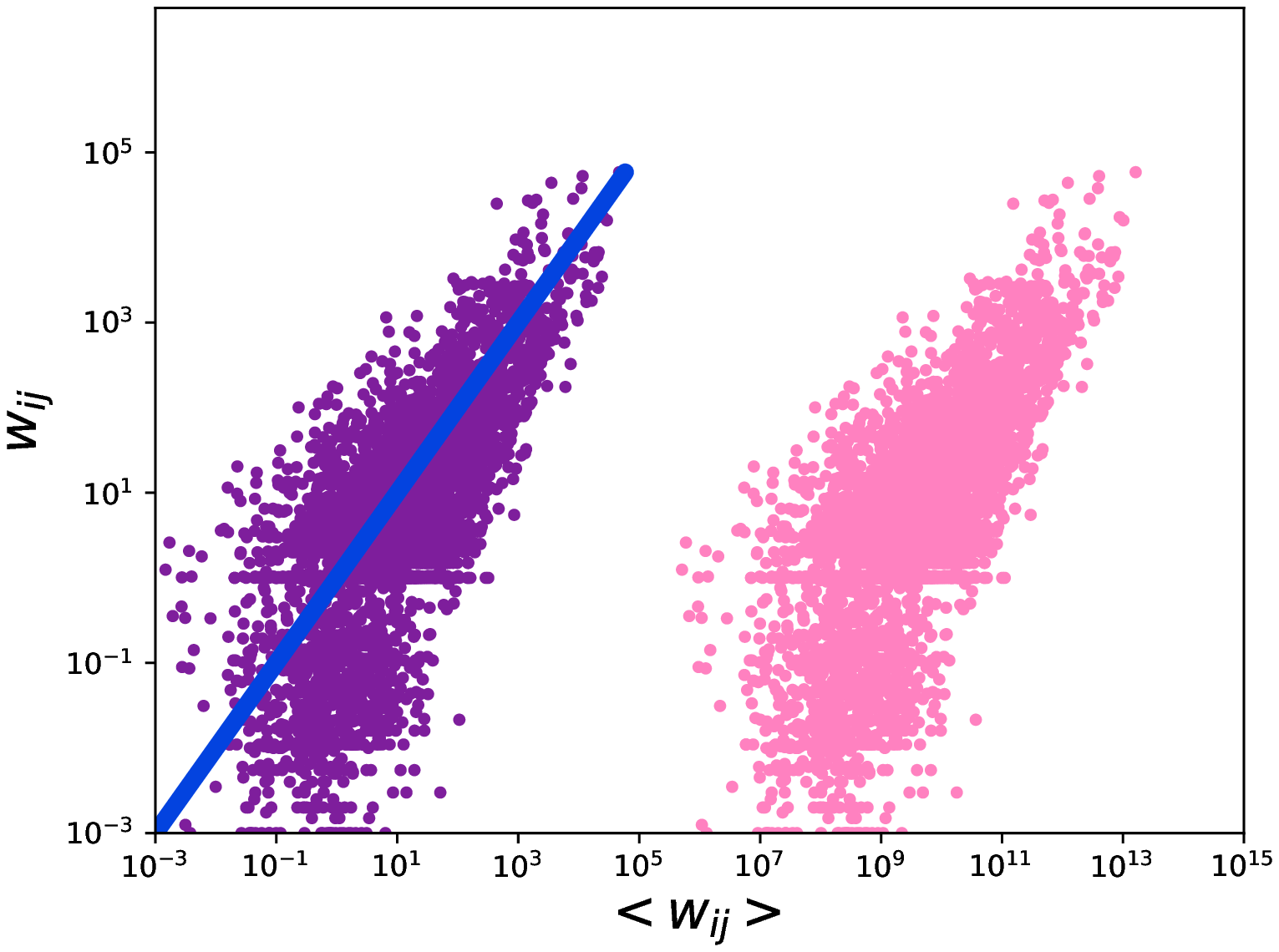}
\label{fig1a}}
\quad
\subfloat[\centering$\rho(\omega_i\omega_j)$ ($\textcolor{xkcdPink}{\bullet}$) VS $\rho(\omega_i\omega_j)d_{ij}^{-1}$ ($\textcolor{xkcdPurple}{\bullet}$)]
{\includegraphics[width=.31\textwidth]{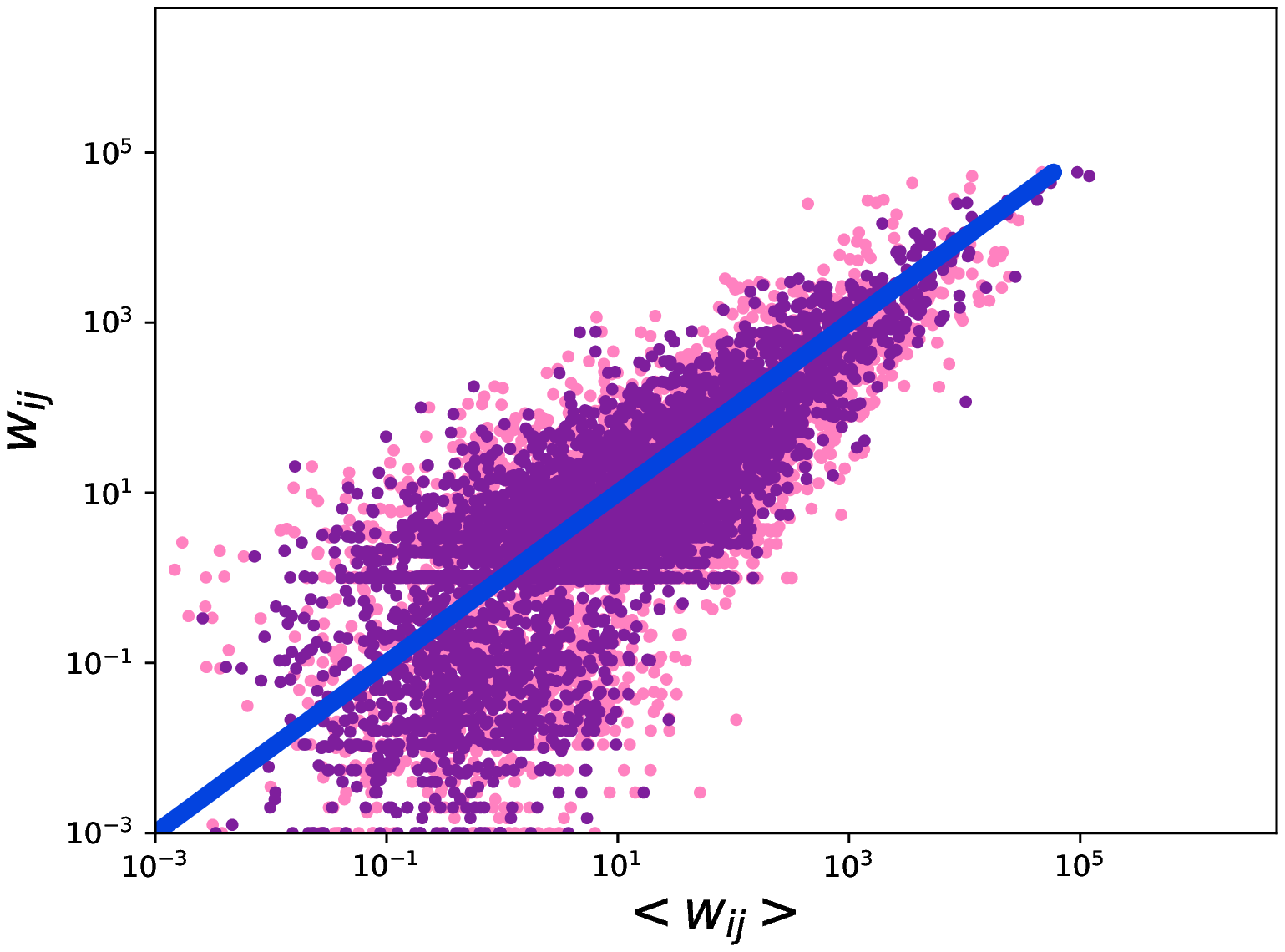}
\label{fig1b}}
\quad
\subfloat[\centering$\rho(\omega_i\omega_j)d_{ij}^{-1}$ ($\textcolor{xkcdPink}{\bullet}$) VS $\rho(\omega_i\omega_j)^\beta d_{ij}^\gamma$ ($\textcolor{xkcdPurple}{\bullet}$)]
{\includegraphics[width=.31\textwidth]{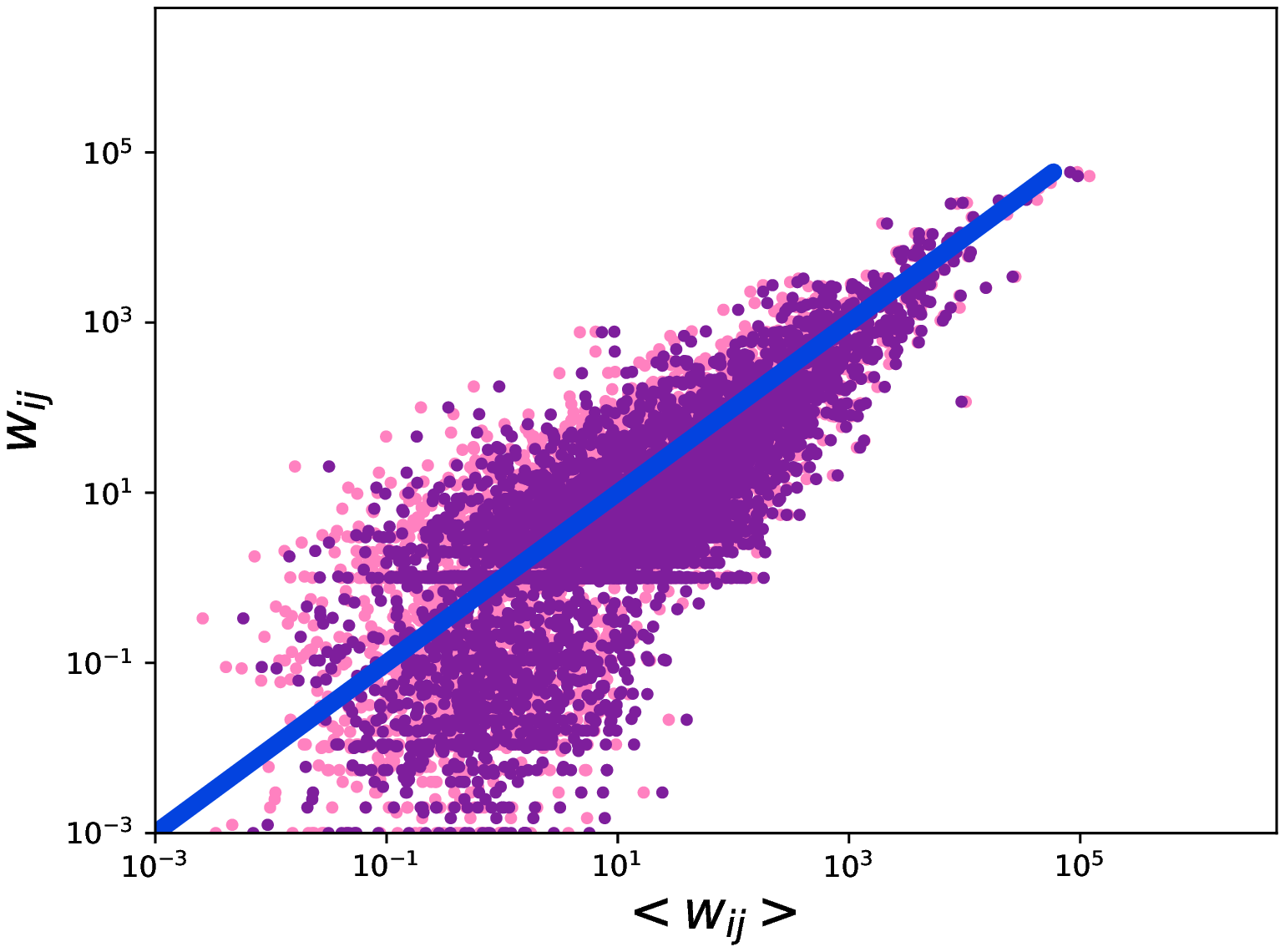}
\label{fig1c}}
\caption{Scatter plot of the entire set of positive WTW weights $w_{ij}$ versus the values predicted by different specifications of the GM: (1) GM with $\rho=\beta=1$ and $\gamma=0$; (2) GM with $\beta=1$, $\gamma=0$ and $\rho$ tuned by requiring that the total weight is reproduced, i.e. $W=\sum_{i<j}w_{ij}=\sum_{i<j}\langle w_{ij}\rangle_\text{GM}=\langle W\rangle_\text{GM}$; (3) GM with $\beta=1$, $\gamma=-1$ and $\rho$ tuned as above; (4) full GM where all parameters are tuned as described in Appendix A. While moving from a parameter-free model to a single-parameter model largely improves the goodness-of-fit, see (a), the net effect of rising the number of explanatory variables is that of decreasing the dispersion around the identity, see (b). Further rising the number parameters, instead, does not add much to the picture provided by the third specification of the GM, see (c). Results refer to the year 2000 of the dataset curated by Gleditsch~\cite{Gleditsch2002}.}
\label{fig1}
\end{figure*}

Our econometric exercise is carried out by comparing the empirical, positive weights of the WTW, in the year 2000, with four different specifications of the following econometric function

\begin{equation}
\langle w_{ij}\rangle_\text{GM}=f(\omega_i,\omega_j,d_{ij}|\underline{\theta})=\rho(\omega_i\omega_j)^{\beta}d_{ij}^{\gamma}
\label{GMexpression}
\end{equation}
where $\omega_i=\frac{\text{GDP}_i}{\overline{\text{GDP}}}$ is the GDP of country $i$ divided by the arithmetic mean of GDPs, $d_{ij}$ is the geographic distance between the capitals of countries $i$ and $j$ and $\underline{\theta}=(\rho,\beta,\gamma)$. The first specification of the model above is characterized by the assignment $\rho=\beta=1$ and $\gamma=0$ and its performance in reproducing the positive weights of the WTW is shown in  Fig.~\ref{fig1}\subref{fig1a}. The overall poor performance of this version of the GM signals the presence of both a \textit{scaling problem} - estimates are positively correlated with observations, only shifted to the right - and of a \textit{dimensional problem} - in fact, pure numbers appear on the right-hand side while the WTW weights are measured in (multiples of) dollars.

A second specification of the GM solves both problems at once. This specification is characterized by the assignment $\beta=1$ and $\gamma=0$ while $\rho$ is treated as a free parameter, tuned by requiring that the total weight is reproduced, i.e. that $W=\sum_{i<j}w_{ij}=\sum_{i<j}\langle w_{ij}\rangle_\text{GM}=\langle W\rangle_\text{GM}$: the fit is, now, much more accurate, as shown in Fig.~\ref{fig1}\subref{fig1b}. The model can be further enriched by adding dyadic factors such as the geographic distances between capitals: some more accuracy in the description of the empirical data is indeed gained, as the reduced dispersion of the cloud of points around the identity witness. Quite remarkably, the picture does not change much if, now, we let the entire set of parameters to be tuned as described in Appendix A (see Fig.~\ref{fig1}\subref{fig1c}).

Although the GM leads to a good prediction of the positive weights, its intrinsic limitation is that it does not allow the topological structure of the WTW to be correctly recovered. In fact, by outputting only positive weights, it induces a trivial, fully-connected structure: upon defining $a_{ij}=\Theta[\langle w_{ij}\rangle_\text{GM}]$, $\forall\:i<j$, it is evident that $a_{ij}=1$, $\forall\:i<j$. In order to overcome such a limitation, one can `refine' the plain gravity model by `dressing' it with a probability distribution capable of accounting for the null entries as well.

In very general terms, we need to define a \textit{statistical network model}, i.e. a set of mathematical relationships between the random variables that are of interest for our network description. Two broad classes of such models can be identified, i.e. the \textit{econometric} ones and the ones rooted into \textit{statistical physics}. In what follows, we will deal with (either econometric or physics-rooted) \textit{discrete} statistical models.

\section{Statistical Network Models - I:\\Econometric Models\label{secIII}}

Let us start with the description of some of the most representative members of the econometric class of models.

\subsection{Poisson Model}

The simplest model in this class prescribes to consider $\langle w_{ij}\rangle_\text{GM}=\rho(\omega_i\omega_j)^{\beta}d_{ij}^{\gamma}$ as the expected value of a Poisson probability mass function:

\begin{equation}
q_{ij}^\text{Pois}(w_{ij})=\frac{z_{ij}^{w_{ij}}e^{-z_{ij}}}{w_{ij}!};
\end{equation}
since $\langle w_{ij}\rangle_\text{Pois}=z_{ij}$, the explanatory power of the GM is retained upon requiring that $\langle w_{ij}\rangle_\text{Pois}=\langle w_{ij}\rangle_\text{GM}$, i.e. by posing

\begin{equation}
z_{ij}=\rho(\omega_i\omega_j)^{\beta}d_{ij}^{\gamma}.
\end{equation}

The expected topology of the network is determined by the expected adjacency matrix entries implied by the model, which is captured by the expression $\langle a_{ij}\rangle_\text{Pois}=p_{ij}^\text{Pois}=1-q_{ij}^\text{Pois}(0)=1-e^{-z_{ij}}$ (see Appendix B for a detailed description of the procedure to estimate the parameters of the Poisson model).

\subsection{Negative Binomial Model}

The main drawback of the Poisson model is that of predicting a variance of the weights that is necessarily equal to their average value. In general, this may be different from what empirical analyses suggest. In order to overcome this problem, econometricians have considered a different probability mass function, namely the negative binomial one with the introduction of an overdispersion\footnote{In the Poisson case, one has that $\sigma^2_\text{Pois}[w_{ij}]=z_{ij}=\langle w_{ij}\rangle_\text{Pois}$; hence, variance cannot be 
adjusted independently from the mean. In the negative binomial case, instead, $\sigma^2_\text{NB}[w_{ij}]=
z_{ij}(1+\alpha z_{ij})=z_{ij}\left(1+\frac{z_{ij}}{m}\right)=\langle w_{ij}\rangle_\text{NB}(1+\alpha \langle w_{ij}\rangle_\text{NB})=\langle w_{ij}\rangle_\text{NB}\left(1+\frac{\langle w_{ij}\rangle_\text{NB}}{m}\right)$ and the variance can be increased to overcome the problem of overestimating the link density.} parameter $\alpha=m^{-1}$~\cite{Hilbe}:

\begin{equation}
q_{ij}^\text{NB}(w_{ij})=\binom{m+w_{ij}-1}{w_{ij}}\left(\frac{1}{1+\alpha z_{ij}}\right)^m\left(\frac{\alpha z_{ij}}{1+\alpha z_{ij}}\right)^{w_{ij}}.
\end{equation}
 
One finds that $\langle w_{ij}\rangle_\text{NB}=m\alpha z_{ij}=z_{ij}$. The requirement that the expected value of the negative binomial distribution coincides with the prediction coming from the GM, i.e. $\langle w_{ij}\rangle_\text{NB}=\langle w_{ij}\rangle_\text{GM}$, can be, again, realized by posing $z_{ij}=\rho(\omega_i\omega_j)^{\beta}d_{ij}^{\gamma}$. Predictions about the topology are, now, carried out via the expression $\langle a_{ij}\rangle_\text{NB}=p_{ij}^\text{NB}=1-q_{ij}^\text{NB}(0)=1-\left(\frac{1}{1+\alpha z_{ij}}\right)^m$ (see Appendix B for a detailed description of the procedure to estimate the parameters of the negative binomial model).

\subsection{Zero-Inflated Poisson Model}

The main drawback of the econometric models above is that of failing in reproducing the link density of the WTW. For instance, the latter equals $c=\frac{2L}{N(N-1)}\simeq0.63$ in year 2000. It turns out that, while the Poisson model overestimates this quantity, the negative binomial one underestimates it, i.e.

\begin{equation}
\langle c\rangle_\text{NB}<c<\langle c\rangle_\text{Pois}
\end{equation}
where $\langle c\rangle_\text{Pois}=\sum_{i<j}p_{ij}^\text{Pois}\simeq0.68$ and $\langle c\rangle_\text{NB}\sum_{i<j}p_{ij}^\text{NB}\simeq0.60$. For this reason, econometricians have defined the so-called \textit{zero-inflated} (ZI) models, i.e. two-step recipes whose general form reads

\begin{eqnarray}
Q(\mathbf{W})&=&\prod_{i<j}q_{ij}(w_{ij})\nonumber\\
&=&\prod_{i<j}p_{ij}^{a_{ij}}(1-p_{ij})^{1-a_{ij}}\cdot q_{ij}(w_{ij}|a_{ij})\nonumber\\
&=&P(\mathbf{A})Q(\mathbf{W}|\mathbf{A})
\end{eqnarray}
a relationship indicating that the probability of the (network represented by the) weighted adjacency matrix $\mathbf{W}$ can be obtained as the product of the probability $P(\mathbf{A})$ of observing the purely binary adjacency matrix $\mathbf{A}$ and the conditional probability $Q(\mathbf{W}|\mathbf{A})$ - where, for consistency, $\mathbf{A}=\Theta[\mathbf{W}]$, i.e. $a_{ij}=\Theta[w_{ij}]$, $\forall\:i<j$, the position $a_{ij}=\Theta[w_{ij}]$ meaning that $a_{ij}=1$ whenever $w_{ij}>0$ and $a_{ij}=0$ if and only if $w_{ij}=0$.\\

The simplest ZI model is the Poisson one, defined by the positions

\begin{eqnarray}
p_{ij}^\text{ZIP}&=&\frac{G_{ij}}{1+G_{ij}}(1-e^{-z_{ij}}),\\
q_{ij}^\text{ZIP}(w_{ij}|a_{ij}=1)&=&
\begin{cases}
\frac{z_{ij}^{w_{ij}}e^{-z_{ij}}}{(1-e^{-z_{ij}})w_{ij}!}, & w_{ij}>0\\
0, & w_{ij}\leq0
\end{cases}.
\end{eqnarray}
Notice that $1-p_{ij}^\text{ZIP}=\frac{1}{1+G_{ij}}+\frac{G_{ij}}{1+G_{ij}}e^{-z_{ij}}$, i.e. the connection between nodes $i$ and $j$ can be missing either because a link is not there (with probability $\frac{1}{1+G_{ij}}$) or because a link is there but has zero weight (with probability $\frac{G_{ij}}{1+G_{ij}}e^{-z_{ij}}$). Consistently, $i$ and $j$ are connected because the weight is not zero (with probability $1-e^{-z_{ij}}$). In order to `dress' the GM, we need to identify some of the parameters of the Poisson model with the usual econometric function. Since

\begin{eqnarray}\label{eqzip}
\langle w_{ij}\rangle_\text{ZIP}&=&p_{ij}^\text{ZIP}\langle w_{ij}|a_{ij}\rangle_\text{ZIP}\nonumber\\
&=&p_{ij}^\text{ZIP}\frac{z_{ij}}{1-e^{-z_{ij}}}=\frac{G_{ij}}{1+G_{ij}}z_{ij},
\end{eqnarray}
we can make the identification $z_{ij}=\rho(\omega_i\omega_j)^{\beta}d_{ij}^{\gamma}$. Upon doing so, we are treating $z_{ij}$ as an `effective' conditional weight: in fact, Eq.~(\ref{eqzip}) can be understood as describing an aleatory experiment that combines a logit with a full Poisson step. According to this interpretation, $z_{ij}$ would represent a Poisson-like expected weight, conditional to the success of the logit step, i.e. $z_{ij}=\frac{\langle w_{ij}\rangle_\text{ZIP}}{p_{ij}^\text{logit}}$, with $p_{ij}^\text{logit}=\frac{G_{ij}}{1+G_{ij}}$.

A second econometric identification is, however, needed: we will proceed by imposing

\begin{equation}
G_{ij}=\delta\omega_i\omega_j
\end{equation}
(see Appendix B for a detailed description of the procedure to estimate the parameters of the ZIP model).

\subsection{Zero-Inflated Negative Binomial Model}

The ZI version of the negative binomial model, instead, is defined by 

\begin{widetext}
\begin{eqnarray}
p_{ij}^\text{ZINB}&=&\frac{G_{ij}}{1+G_{ij}}(1-\tau_{ij}),\\
q_{ij}^\text{ZINB}(w_{ij}|a_{ij}=1)&=&
\begin{cases}
\binom{m+w_{ij}-1}{w_{ij}}\left(\frac{1}{1-\tau_{ij}}\right)\left(\frac{1}{1+\alpha z_{ij}}\right)^m\left(\frac{\alpha z_{ij}}{1+\alpha z_{ij}}\right)^{w_{ij}}, & w_{ij}>0\\
0, & w_{ij}\leq0
\end{cases}
\end{eqnarray}
\end{widetext}
where, as the for the plain negative binomial model, $\alpha=m^{-1}$ and $\tau_{ij}=\left(\frac{1}{1+\alpha z_{ij}}\right)^m$. Moreover, as for the zero-inflated Poisson (ZIP) model, $1-p_{ij}^\text{ZINB}=\frac{1}{1+G_{ij}}+\frac{G_{ij}}{1+G_{ij}}\tau_{ij}$, i.e. the connection between nodes $i$ and $j$ can be missing either because a link is not there (with probability $\frac{1}{1+G_{ij}}$) or because a link is there but has zero weight (with probability $\frac{G_{ij}}{1+G_{ij}}\tau_{ij}$); consistently, $i$ and $j$ are connected because the weight is not zero (with probability $1-\tau_{ij}$). In order to `dress' the GM, we need to identify some of the parameters of the negative binomial model with the usual econometric function. Upon considering that 

\begin{eqnarray}
\langle w_{ij}\rangle_\text{ZINB}&=&p_{ij}^\text{ZINB}\langle w_{ij}|a_{ij}\rangle_\text{ZINB}\nonumber\\
&=&p_{ij}^\text{ZINB}\frac{z_{ij}}{1-\tau_{ij}}=\frac{G_{ij}}{1+G_{ij}}z_{ij},
\end{eqnarray}
we can make the identification $z_{ij}=\rho(\omega_i\omega_j)^{\beta}d_{ij}^{\gamma}$ and $G_{ij}=\delta\omega_i\omega_j$.

As for the ZIP case, we are treating $z_{ij}$ as a negative binomial-like expected weight, conditional to the success of a logit step, i.e. $z_{ij}=\frac{\langle w_{ij}\rangle_\text{ZINB}}{p_{ij}^\text{logit}}$, with $p_{ij}^\text{logit}=\frac{G_{ij}}{1+G_{ij}}$ (see Appendix B for a detailed description of the procedure to estimate the parameters of the ZINB model).\\

Let us notice that, while the ZIP model provides a better estimation of the link density than the Poisson model, the ZINB and the negative binomial ones basically perform in the same way. In fact, 

\begin{equation}
\langle c\rangle_\text{ZINB}<c\simeq\langle c\rangle_\text{ZIP}
\end{equation}
since $c=\frac{2L}{N(N-1)}\simeq0.63$, $\langle c\rangle_\text{ZIP}=\sum_{i<j}p_{ij}^\text{ZIP}\simeq0.63$ and $\langle c\rangle_\text{ZINB}=\sum_{i<j}p_{ij}^\text{ZINB}\simeq0.60$, a result suggesting that both variants of the negative binomial model will perform poorly in reproducing the binary properties of the WTW.

\section{Statistical Network Models - II:\\Maximum-Entropy Models\label{secIV}}

The members of the second class of network models are the ones defined within the framework of traditional statistical mechanics. All of them can be derived by performing a constrained maximization of Shannon entropy~\cite{Squartini2015b} where the constraints represent the available information about the system at hand.\\

The simplest, yet non trivial, ME model that can be considered comes from the maximization of the binary Shannon functional

\begin{equation}
S=-\sum_\mathbf{A}P(\mathbf{A})\ln P(\mathbf{A})
\end{equation}
constrained to reproduce the entire degree sequence, $\{k_i(\mathbf{A})\}_{i=1}^N$, of the network. This model is known under the name of Undirected Binary Configuration Model (UBCM) and has been shown to accurately reproduce many (binary) properties of a wide spectrum of real-world systems~\cite{Squartini2011a}.

The UBCM is described by the probability mass function

\begin{equation}
P(\mathbf{A})=\prod_{i<j}p_{ij}^{a_{ij}}(1-p_{ij})^{1-a_{ij}},
\end{equation}
which is factorized into the product of Bernoulli probability mass functions (one for each pair of nodes) with 

\begin{equation}
p_{ij}^\text{UBCM}=\frac{x_ix_j}{1+x_ix_j}
\end{equation}
(where $x_i$ is the Lagrange multiplier controlling for the degree of node $i$). Importantly, the logit model admitting the presence of a single global constant can be derived from entropy maximization upon re-parametrizing the Lagrange multipliers of the UBCM and imposing the total number of links as the only constraint~\cite{Squartini2018}). The identification $x_i\equiv\sqrt{\delta}\omega_i$, in fact, leads to

\begin{equation}
p_{ij}^\text{logit}=\frac{G_{ij}}{1+G_{ij}}=\frac{\delta\omega_i\omega_j}{1+\delta\omega_i\omega_j}.
\end{equation}
Although the functional form above is not the most general one (for instance, dyadic factors such as geographic distances could be added as well), it is the form we will adopt in what follows.

In the network literature, the logit model (in its formulation above) has been popularized~\cite{Garlaschelli2004} as one particular case of the so-called fitness model~\cite{Caldarelli2002} and as the so-called density-corrected Gravity Model (dcGM)~\cite{Cimini2015} and has been proven to perform remarkably well for the task of reconstructing the topology of networks from partial information~\cite{Squartini2018}.\\

Since we are interested in reproducing the structural properties of weighted networks, we need to complement the purely binary step above with a recipe for reconstructing weights. The entropy-based framework handles such a requirement via the maximization of conditional Shannon functionals allowing the specification of $P(\mathbf{A})$ to be disentangled from that of $Q(\mathbf{W}|\mathbf{A})$~\cite{Parisi2020}.

When discrete weighted models are considered, a useful quantity is the conditional Shannon entropy 

\begin{equation}
S(\mathscr{W}|\mathscr{A})=-\sum_{\mathbf{A}\in\mathbb{A}}P(\mathbf{A})\sum_{\mathbb{W}_\mathbf{A}}Q(\mathbf{W}|\mathbf{A})\log Q(\mathbf{W}|\mathbf{A})
\end{equation}
where the first sum runs over all binary configurations within the ensemble $\mathbb{A}$ and the second sum runs over all weighted configurations that are compatible with each specific binary structure represented by the adjacency matrix $\mathbf{A}$, i.e. such that $\mathbb{W}_\mathbf{A}=\{\mathbf{W}:\Theta[\mathbf{W}]=\mathbf{A}\}$.

Conditional maximization proceeds by specifying a set of weighted constraints that, in the discrete case, reads

\begin{eqnarray}
1&=&\sum_{\mathbb{W}_\mathbf{A}}P(\mathbf{W}|\mathbf{A}),\:\forall\:\mathbf{A}\in\mathbb{A}\\
\langle C_\alpha\rangle&=&\sum_{\mathbf{A}\in\mathbb{A}}P(\mathbf{A})\sum_{\mathbb{W}_\mathbf{A}}Q(\mathbf{W}|\mathbf{A})C_\alpha(\mathbf{W}),\:\forall\:\alpha
\end{eqnarray}
the first condition ensuring the normalization of the conditional probability mass function and the vector $\{C_\alpha(\mathbf{W})\}$ representing the `proper' set of weighted constraints. Differentiating the corresponding Lagrangean functional with respect to $Q(\mathbf{W}|\mathbf{A})$ and equating the result to zero leads to 

\begin{equation}
Q(\mathbf{W}|\mathbf{A})=
\begin{cases}
\frac{e^{-H(\mathbf{W})}}{Z_\mathbf{A}}, & \:\mathbf{W}\in\mathbb{W}_\mathbf{A}\\
0, & \:\mathbf{W}\notin\mathbb{W}_\mathbf{A}
\end{cases}
\end{equation}
where $H(\mathbf{W})=\sum_\alpha\psi_\alpha C_\alpha$ is the so-called Hamiltonian, listing the constrained, weighted quantities, and $Z_\mathbf{A}=\sum_{\mathbb{W}_\mathbf{A}}e^{-H(\mathbf{W})}$ is the partition function for fixed $\mathbf{A}$. The explicit functional form of $Q(\mathbf{W}|\mathbf{A})$ can be obtained only once the functional form of the constraints has been specified as well.\\

To this aim, let us consider the Hamiltonian

\begin{equation}
H(\mathbf{W})=\sum_{i<j}\phi_{ij}w_{ij},
\end{equation}
where weights are modelled as non-negative integer variables, i.e. $w_{ij}\in\mathbb{N}$, $\forall\:i<j$. This choice induces a conditional probability mass function reading

\begin{equation}
Q(\mathbf{W}|\mathbf{A})=\prod_{i<j}q_{ij}(w_{ij}|a_{ij})=\prod_{i<j}y_{ij}^{w_{ij}-a_{ij}}(1-y_{ij})^{a_{ij}}
\end{equation}
(with $e^{-\psi_{ij}}=e^{-\phi_{ij}}=y_{ij}$). Let us, now, turn the model above into a proper econometric one. To this aim, let us proceed by analogy. All zero-inflated econometric recipes identify $z_{ij}$ with a conditional expected weight, a prescription that in our case, would translate into its identification with $\langle w_{ij}|a_{ij}\rangle=\frac{1}{1-y_{ij}}$. This choice, however, would lead to an inconsistency, since $z_{ij}>0$ is a positive real number while $\langle w_{ij}|a_{ij}\rangle$ must necessarily exceed 1, as it represents the expected weight conditional to the existence of a connection. An alternative, consistent econometric identification is

\begin{equation}\label{econ1}
\frac{1}{1-y_{ij}}=1+z_{ij}
\end{equation}
which, in turn, induces a conditional probability mass function 

\begin{eqnarray}
Q(\mathbf{W}|\mathbf{A})&=&\prod_{i<j}q_{ij}(w_{ij}|a_{ij})\nonumber\\
&=&\prod_{i<j}\left(\frac{z_{ij}}{1+z_{ij}}\right)^{w_{ij}-a_{ij}}\left(\frac{1}{1+z_{ij}}\right)^{a_{ij}}.
\end{eqnarray}

Models of the kind are known as hurdle models: quite remarkably, entropy maximization allows us to recover them in a fully principled way, i.e. by eliminating the (otherwise unavoidable) ambiguity that accompanies the choice of the distribution (supposedly) describing the positive values of an economic system.

The hurdle-geometric model derived above, however, suffers from a number of limitations, the most relevant of which is that of failing in reproducing basic network quantities such as the WTW total weight. As an illustrative example, while $W=\sum_{i<j}w_{ij}\simeq 10^9$, in the year 2000, we find that $\langle W\rangle_\text{h-g}\simeq 10^5$. In order to overcome such a limitation, we have considered the conditional probability mass function induced by the Hamiltonian

\begin{equation}\label{hamw}
H(\mathbf{W})=\sum_{i<j}(\phi_0+\phi_{ij})w_{ij}=\phi_0 W+\sum_{i<j}\phi_{ij}w_{ij}.
\end{equation}
Identifying $e^{-\phi_0}=y_0$ and $e^{-\phi_{ij}}=y_{ij}=\frac{z_{ij}}{1+z_{ij}}$ (see Eq.~\eqref{econ1}), we arrive at the modified econometric model

\begin{equation}\label{condw}
q_{ij}(w_{ij}|a_{ij})=\left(\frac{y_0z_{ij}}{1+z_{ij}}\right)^{w_{ij}-a_{ij}}\left(\frac{1+z_{ij}-y_0z_{ij}}{1+z_{ij}}\right)^{a_{ij}}.
\end{equation}

We are now ready to fully specify the suite of discrete entropy-models that we will compare with the aforementioned, purely econometric ones. To this aim, we need to fully specify the functional form

\begin{eqnarray}\label{full1}
Q(\mathbf{W})&=&\prod_{i<j}q_{ij}(w_{ij})\nonumber\\
&=&\prod_{i<j}p_{ij}^{a_{ij}}(1-p_{ij})^{1-a_{ij}}\cdot q_{ij}(w_{ij}|a_{ij})\nonumber\\
&=&P(\mathbf{A})Q(\mathbf{W}|\mathbf{A});
\end{eqnarray}
the two most obvious choices are represented by the models 

\begin{eqnarray}
Q_\text{TSF}(\mathbf{W})&=&P_\text{logit}(\mathbf{A})Q(\mathbf{W}|\mathbf{A})\nonumber\\
&=&\prod_{i<j}(p_{ij}^\text{logit})^{a_{ij}}(1-p_{ij}^\text{logit})^{1-a_{ij}}\cdot q_{ij}(w_{ij}|a_{ij})\nonumber\\
\end{eqnarray}
and

\begin{eqnarray}
Q_\text{TS}(\mathbf{W})&=&P_\text{UBCM}(\mathbf{A})Q(\mathbf{W}|\mathbf{A})\nonumber\\
&=&\prod_{i<j}(p_{ij}^\text{UBCM})^{a_{ij}}(1-p_{ij}^\text{UBCM})^{1-a_{ij}}\cdot q_{ij}(w_{ij}|a_{ij})\nonumber\\
\end{eqnarray}
that combine the weighted, conditional step induced by the Hamiltonian defined in Eq.~(\ref{hamw}) with the purely binary logit model and with the Undirected Binary Configuration Model, respectively. The acronyms stand for `two-step fitness' model and `two-step' model and recall the names originally used to define them~\cite{Garlaschelli2005,Almog2017}.\\

Less trivial choices are represented by models whose both binary and weighted portions are jointly determined by the constraints. They can all be recovered as specifications of the generic Hamiltonian

\begin{eqnarray}
H(\mathbf{W})&=&\sum_{i<j}[\theta_{ij}a_{ij}+(\phi_0+\phi_{ij})w_{ij}]\nonumber\\
&=&\sum_{i<j}\theta_{ij}a_{ij}+\phi_0 W+\sum_{i<j}\phi_{ij}w_{ij};
\end{eqnarray}
in what follows, we will consider two different instances of such a function, defined by the choices $\theta_{ij}=\theta_0$ and $\theta_{ij}=\theta_i+\theta_j$. In other words, while we let the weighted parts of these models coincide and read as in Eq.~(\ref{condw}), we allow for the binary part to vary, either constraining the total number of links, $L$, or the entire degree sequence $\{k_i(\mathbf{A})\}_{i=1}^N$. In the first case, our Hamiltonian reads

\begin{equation}
H_{(1)}(\mathbf{W})=\theta_0L+\phi_0W+\sum_{i<j}\phi_{ij}w_{ij}
\end{equation}
and instances the model in Eq.~(\ref{full1}) with

\begin{eqnarray}
p_{ij}^{(1)}&=&\frac{xy_0z_{ij}}{1+z_{ij}-y_0z_{ij}+xy_0z_{ij}}\\ q_{ij}^{(1)}(w_{ij}|a_{ij})&=&\left(\frac{y_0z_{ij}}{1+z_{ij}}\right)^{w_{ij}-a_{ij}}\left(\frac{1+z_{ij}-y_0z_{ij}}{1+z_{ij}}\right)^{a_{ij}}\nonumber\\
\end{eqnarray}
(having posed $e^{-\theta_0}=x$, $e^{-\phi_0}=y_0$ and $e^{-\phi_{ij}}=y_{ij}=\frac{z_{ij}}{1+z_{ij}}$); in the second case, it reads

\begin{equation}
H_{(2)}(\mathbf{W})=\sum_i\theta_ik_i+\psi_0W+\sum_{i<j}\psi_{ij}w_{ij}
\end{equation}
and instances the model in Eq.~(\ref{full1}) with

\begin{eqnarray}
p_{ij}^{(2)}&=&\frac{x_ix_jy_0z_{ij}}{1+z_{ij}-y_0z_{ij}+x_ix_jy_0z_{ij}}\\ q_{ij}^{(2)}(w_{ij}|a_{ij})&=&\left(\frac{y_0z_{ij}}{1+z_{ij}}\right)^{w_{ij}-a_{ij}}\left(\frac{1+z_{ij}-y_0z_{ij}}{1+z_{ij}}\right)^{a_{ij}}\nonumber\\
\end{eqnarray}
(having posed $e^{-\theta_i}=x_i$, $e^{-\phi_0}=y_0$ and $e^{-\phi_{ij}}=y_{ij}=\frac{z_{ij}}{1+z_{ij}}$). Appendix C provides a detailed description of the procedure we have adopted to estimate the parameters entering into the definition of our basket of discrete ME models.\\

\begin{figure*}[t!]
\subfloat[Comparison between the empirical degree distribution and the ones predicted by the econometric models]{\includegraphics[width=.31\textwidth]
{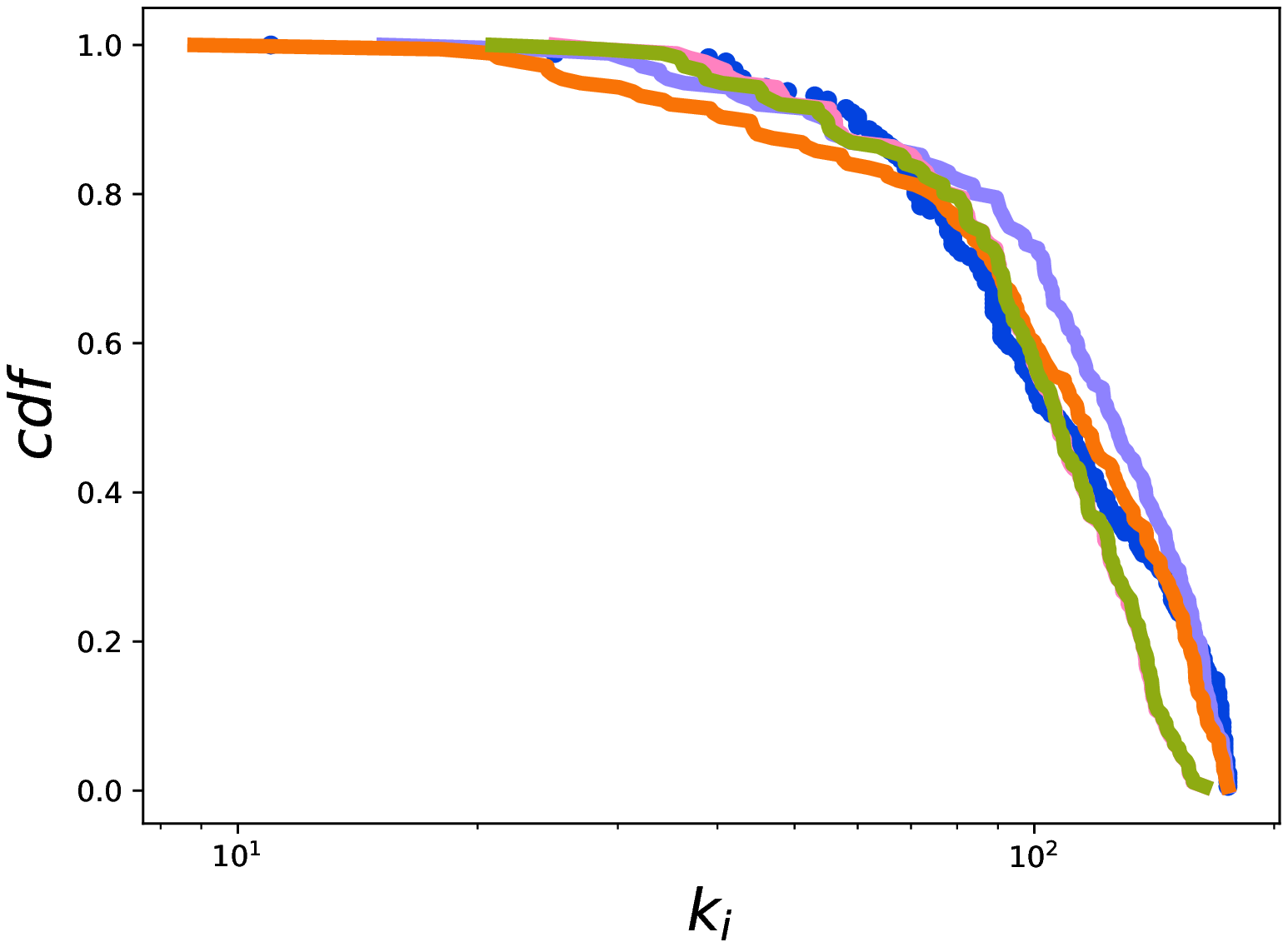}
\label{fig2a}}
\quad
\subfloat[Comparison between the empirical ANND values and the ones predicted by the econometric models]{\includegraphics[width=.31\textwidth]{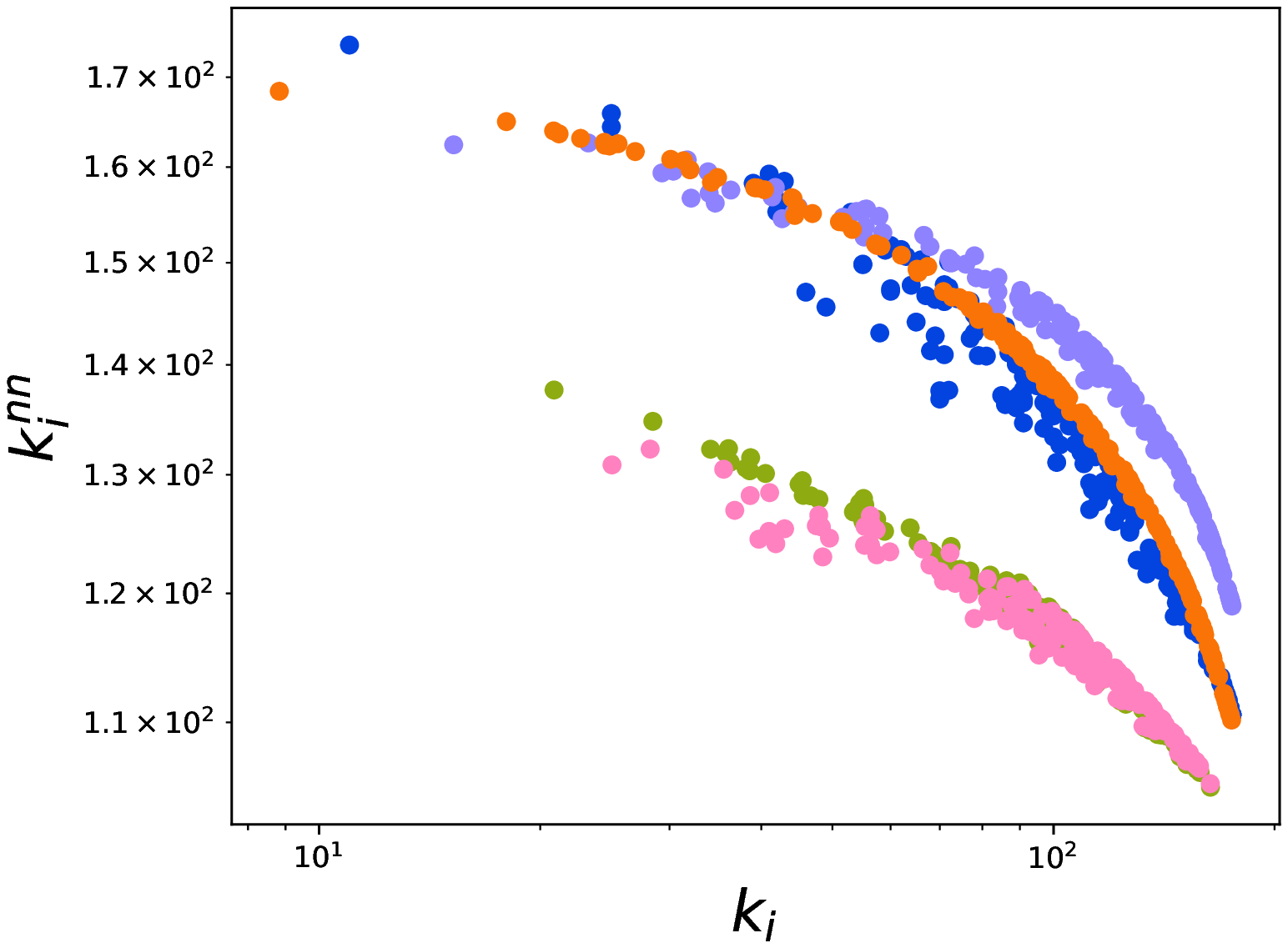}\label{fig2b}}
\quad
\subfloat[Comparison between the empirical BCC values and the ones predicted by the econometric models]{\includegraphics[width=.31\textwidth]{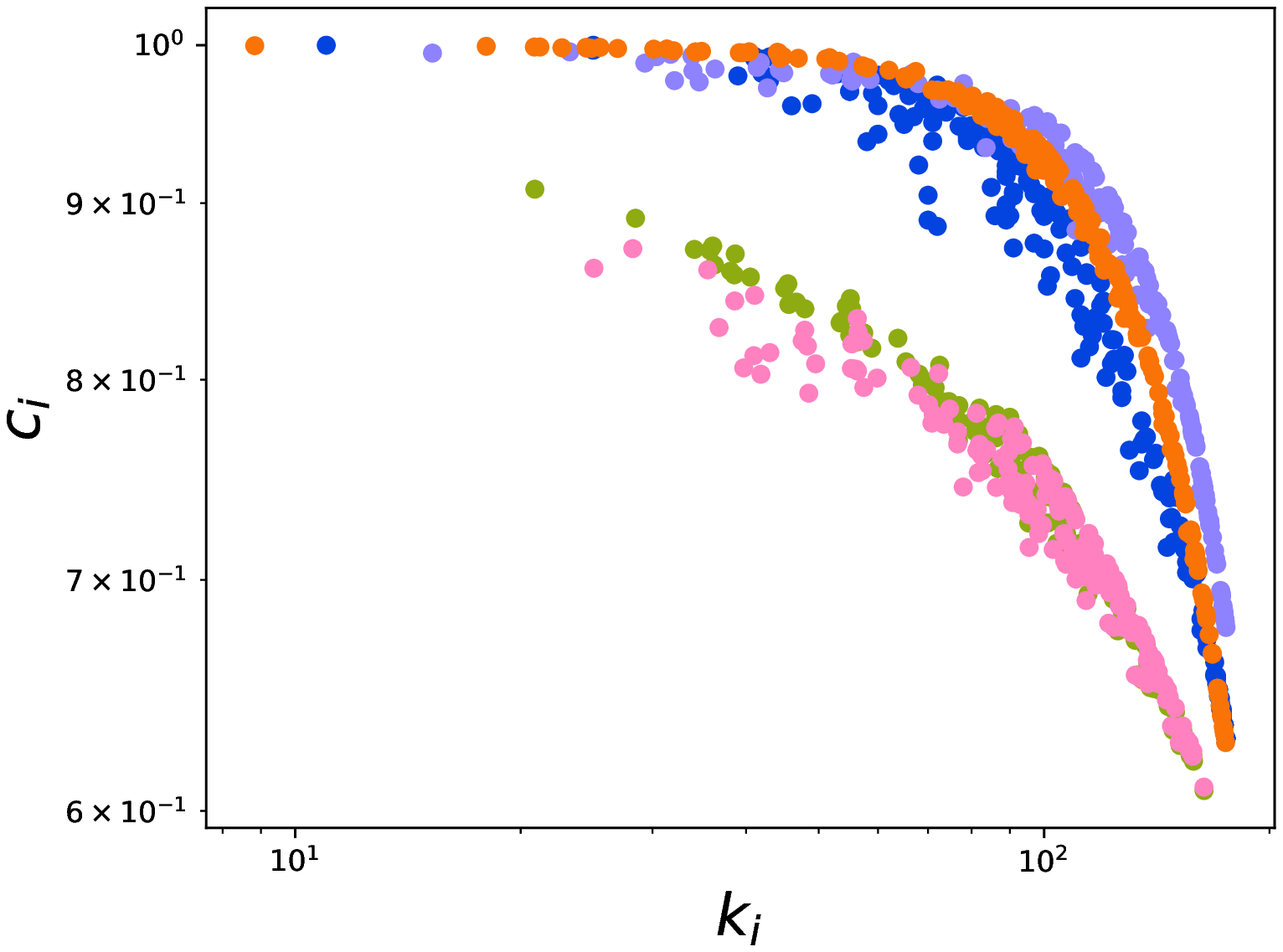}\label{fig2c}}\\
\subfloat[Comparison between the empirical degree distribution and the ones predicted by the ME models]{\includegraphics[width=.31\textwidth]{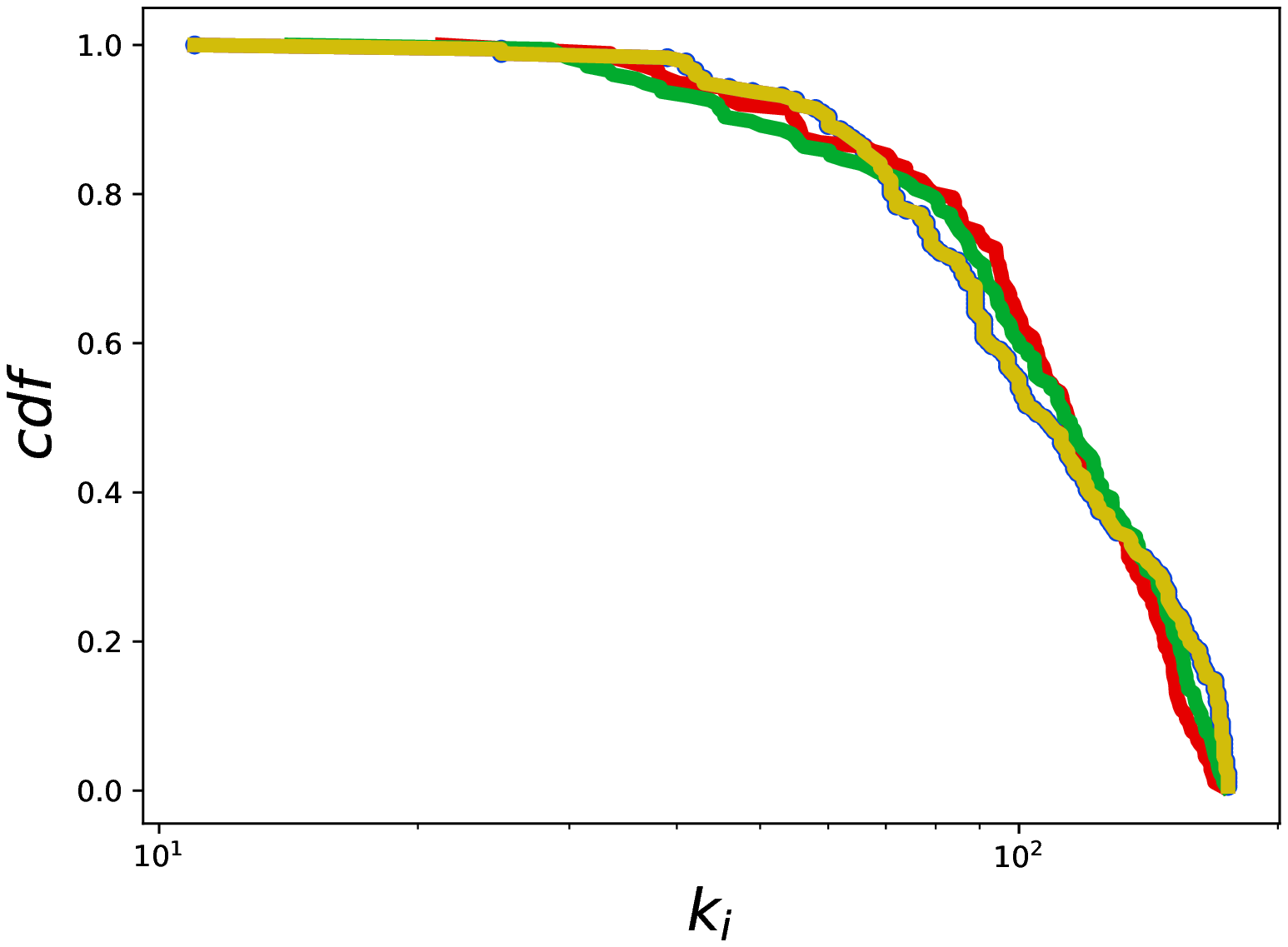}\label{fig2d}}
\quad
\subfloat[Comparison between the empirical ANND values and the ones predicted by the ME models]{\includegraphics[width=.31\textwidth]{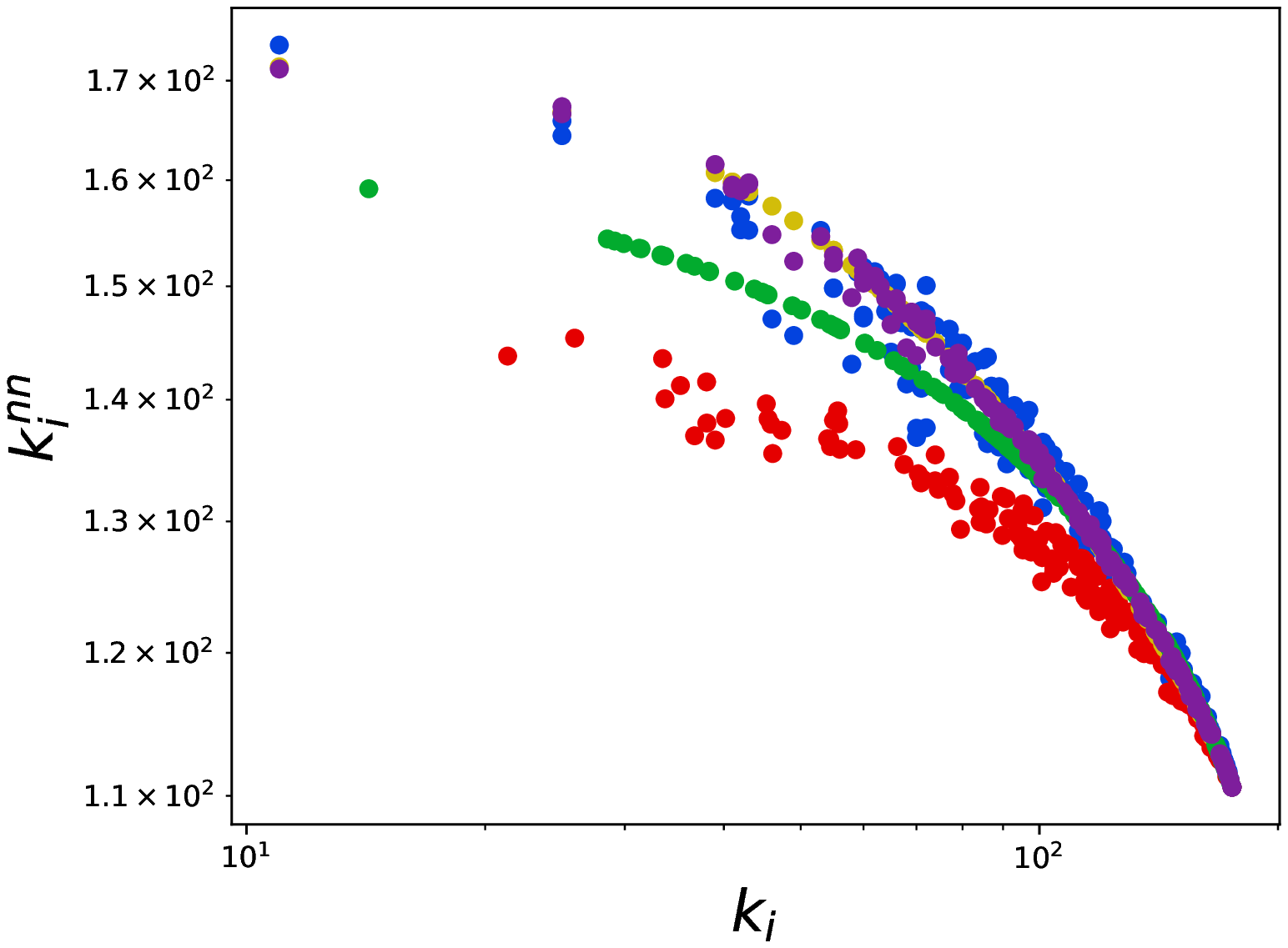}\label{fig2e}}
\quad
\subfloat[Comparison between the empirical BCC values and the ones predicted by the ME models]{\includegraphics[width=.31\textwidth]{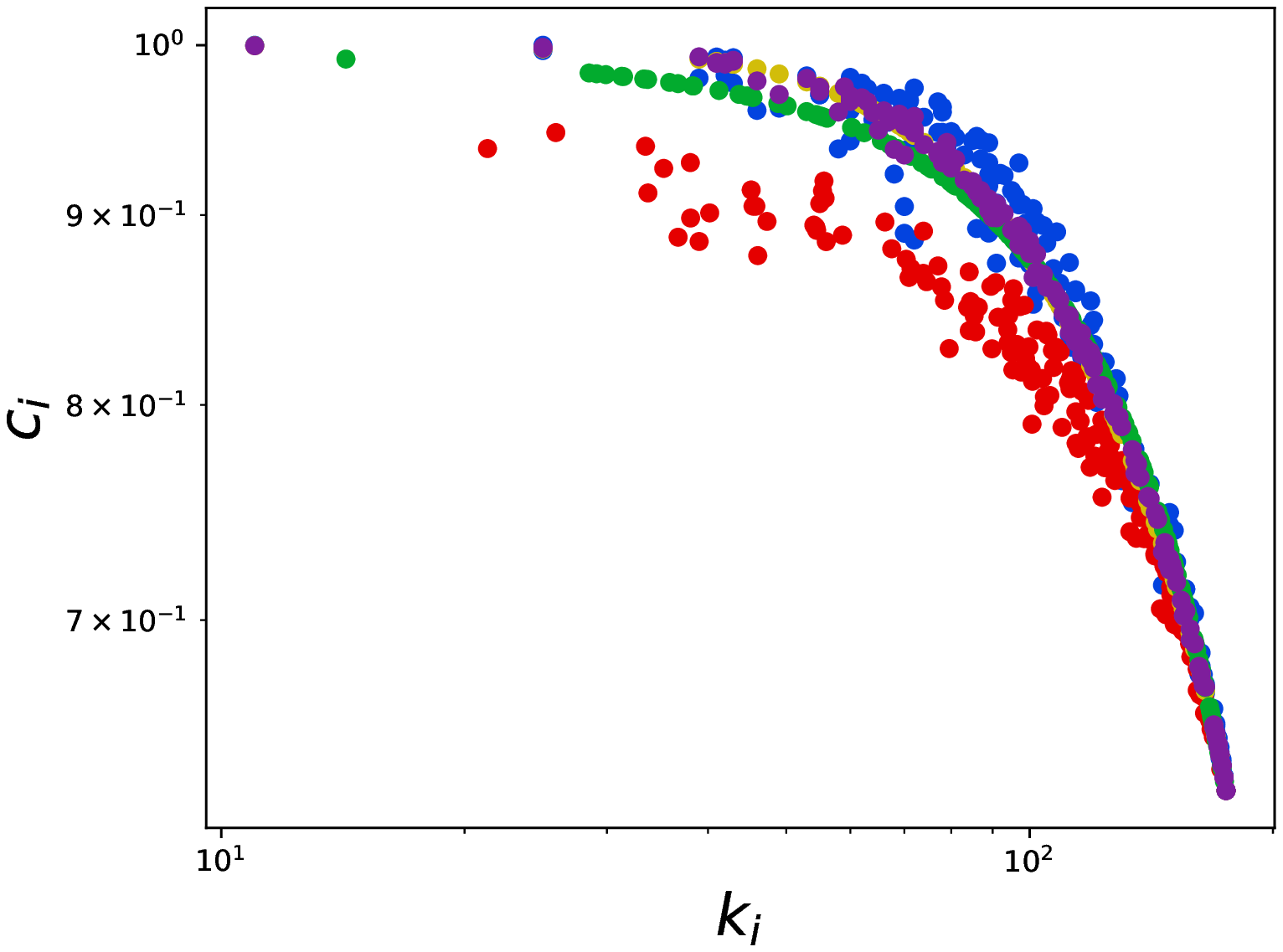}\label{fig2f}}
\caption{Performance of econometric models versus the performance of  ME models in reproducing: (a), (d) the degrees; (b), (e) the ANND; (c), (f) the BCC. Empirical points are indicated as $\textcolor{xkcdBlue}{\bullet}$. Econometric and ME models are indicated as follows: $\textcolor{xkcdPeriwinkle}{\bullet}$ - Poisson; $\textcolor{xkcdPink}{\bullet}$ - Negative Binomial; $\textcolor{xkcdOrange}{\bullet}$ - ZIP; $\textcolor{xkcdPeaGreen}{\bullet}$ - ZINB; $\textcolor{xkcdRed}{\bullet}$ - $H_{(1)}$; $\textcolor{xkcdPurple}{\bullet}$ - $H_{(2)}$; $\textcolor{xkcdMustardYellow}{\bullet}$ - TS; $\textcolor{xkcdKellyGreen}{\bullet}$ - TSF. Results refer to the year 2000 of the dataset curated by Gleditsch~\cite{Gleditsch2002}.}
\label{fig2}
\end{figure*}

So far, we have turned entropy-based models into econometric ones via a suitable econometric transformation of the Lagrange multipliers defining the proper `physical' models. The entropy-based formalism, however, also offers the opportunity to define statistical models in a fully data-driven fashion. To this aim, let us consider the Hamiltonian

\begin{eqnarray}
H_\text{UECM}(\mathbf{W})&=&\sum_i[\theta_ik_i+\phi_is_i]\nonumber\\
&=&\sum_{i<j}[(\theta_i+\theta_j)a_{ij}+(\phi_i+\phi_j)w_{ij}]
\end{eqnarray}
that constrains both degrees and strengths. The model induced by the latter ones is called Undirected Enhanced Configuration Model (UECM) and represents the best-performing one for the task of network reconstruction in presence of full information about the constraints~\cite{Mastrandrea2014,Garlaschelli2009}.\\

Remarkably, all models considered in the previous Section can be compactly derived by combining a logit-like probability mass function describing the binary network structure with the conditional expression defined in Eq.~(\ref{condw}). To prove this, it is enough to notice that all the Bernoulli-like probability mass functions characterizing our model can be compactly rewritten as

\begin{equation}
p_{ij}=\frac{x_i'x_j'}{1+x_i'x_j'}
\end{equation}
where

\begin{eqnarray}
(x_i'x_j')^\text{logit}&=&\delta\omega_i\omega_j,\\
(x_i'x_j')^\text{UBCM}&=&x_ix_j,\\
(x_i'x_j')^{(1)}&=&\frac{xy_0z_{ij}}{1+z_{ij}-y_0z_{ij}},\\
(x_i'x_j')^{(2)}&=&\frac{x_ix_jy_0z_{ij}}{1+z_{ij}-y_0z_{ij}},\\
(x_i'x_j')^\text{UECM}&=&\frac{x_ix_jy_iy_j}{1-y_iy_j}.
\end{eqnarray}

\begin{figure*}[t!]
\subfloat[Comparison between the empirical strength distribution and the ones predicted by the econometric models]{\includegraphics[width=.31\textwidth]{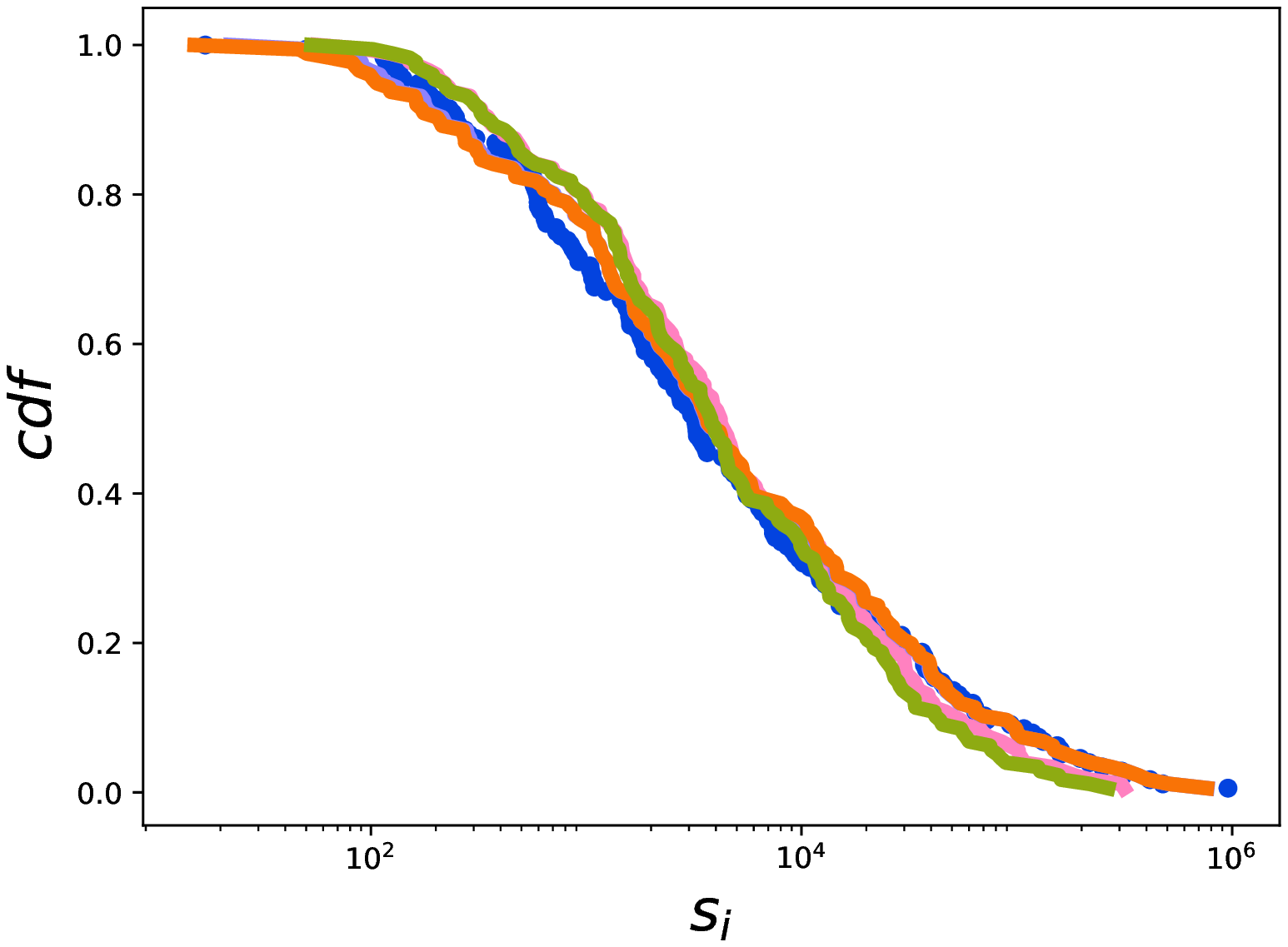}\label{fig3a}}
\quad
\subfloat[Comparison between the empirical ANNS values and the ones predicted by the econometric models]{\includegraphics[width=.31\textwidth]{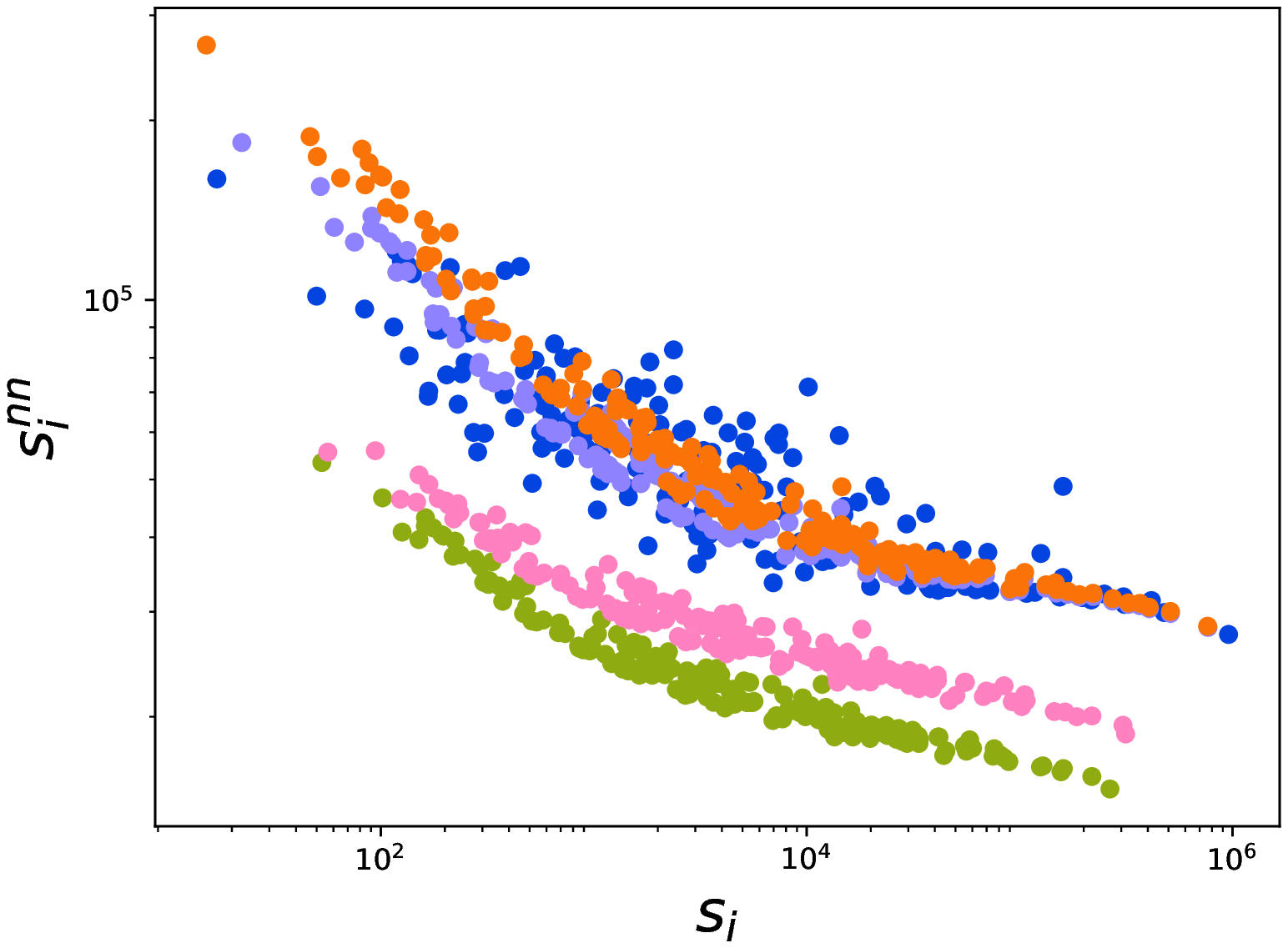}\label{fig3b}}
\quad
\subfloat[Comparison between the empirical WCC values and the ones predicted by the econometric models]{\includegraphics[width=.31\textwidth]{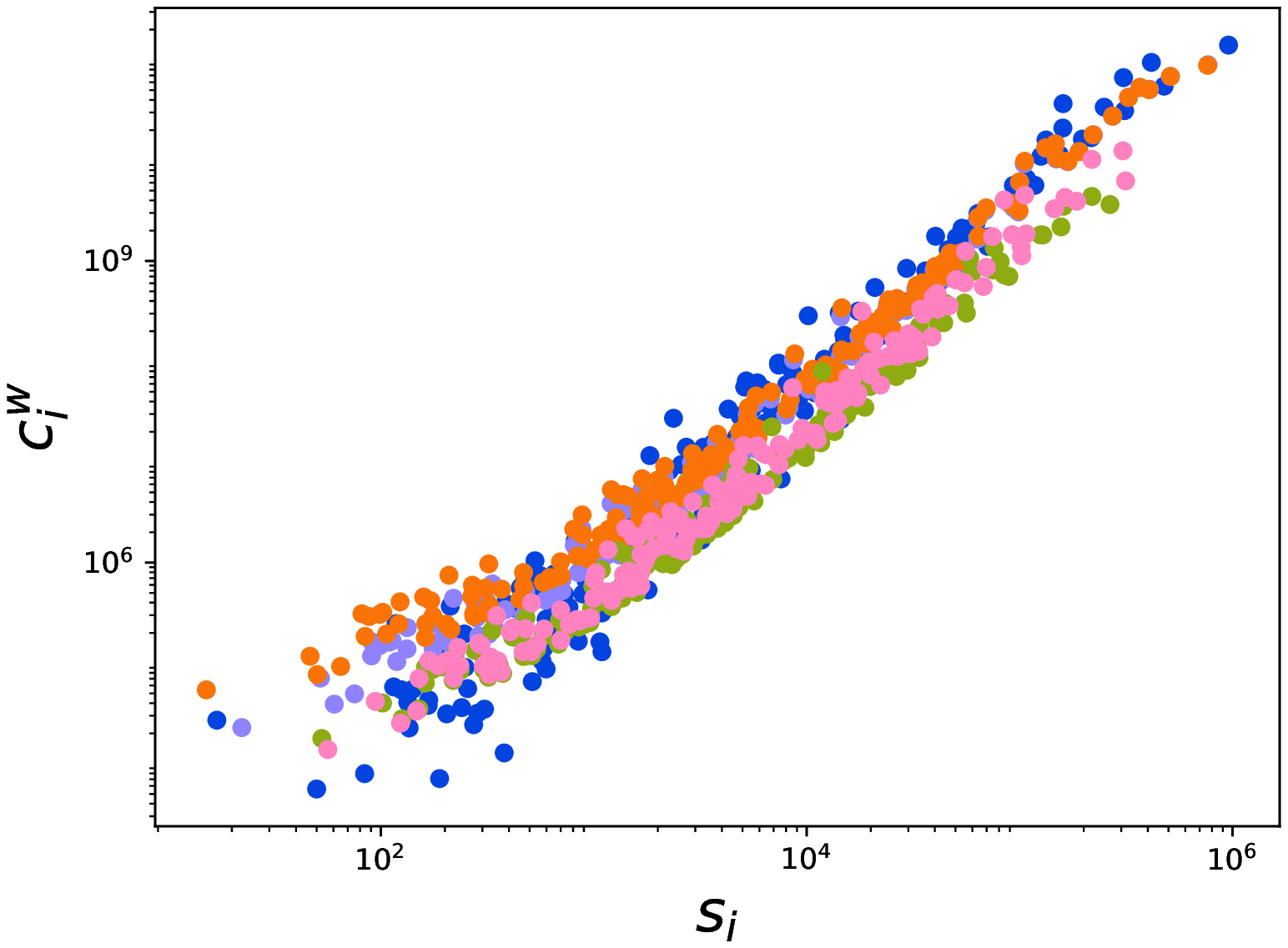}\label{fig3c}}\\
\subfloat[Comparison between the empirical strength distribution and the ones predicted by the ME models]{\includegraphics[width=.31\textwidth]{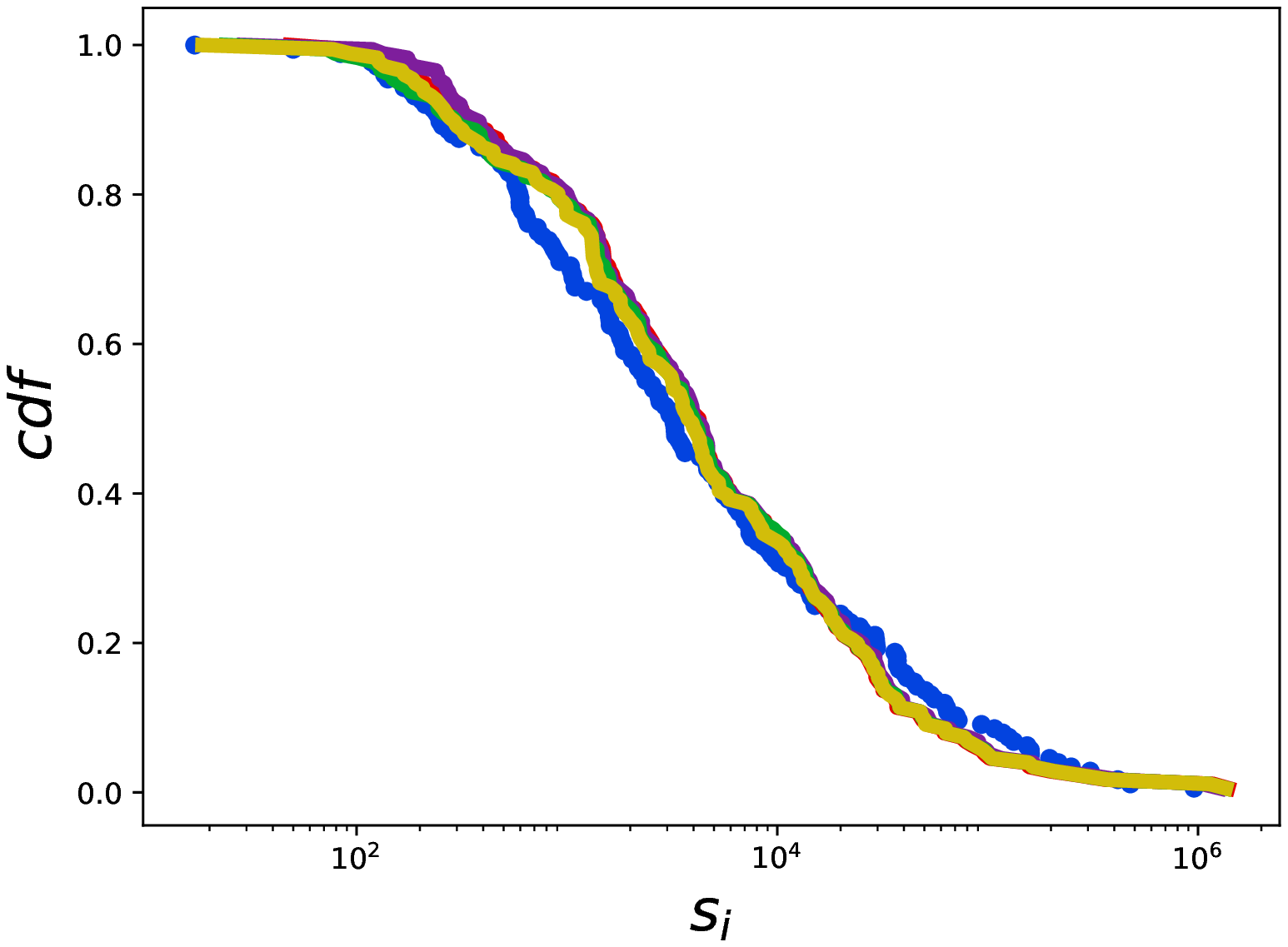}\label{fig3d}}
\quad
\subfloat[Comparison between the empirical ANNS values and the ones predicted by the ME models]{\includegraphics[width=.31\textwidth]{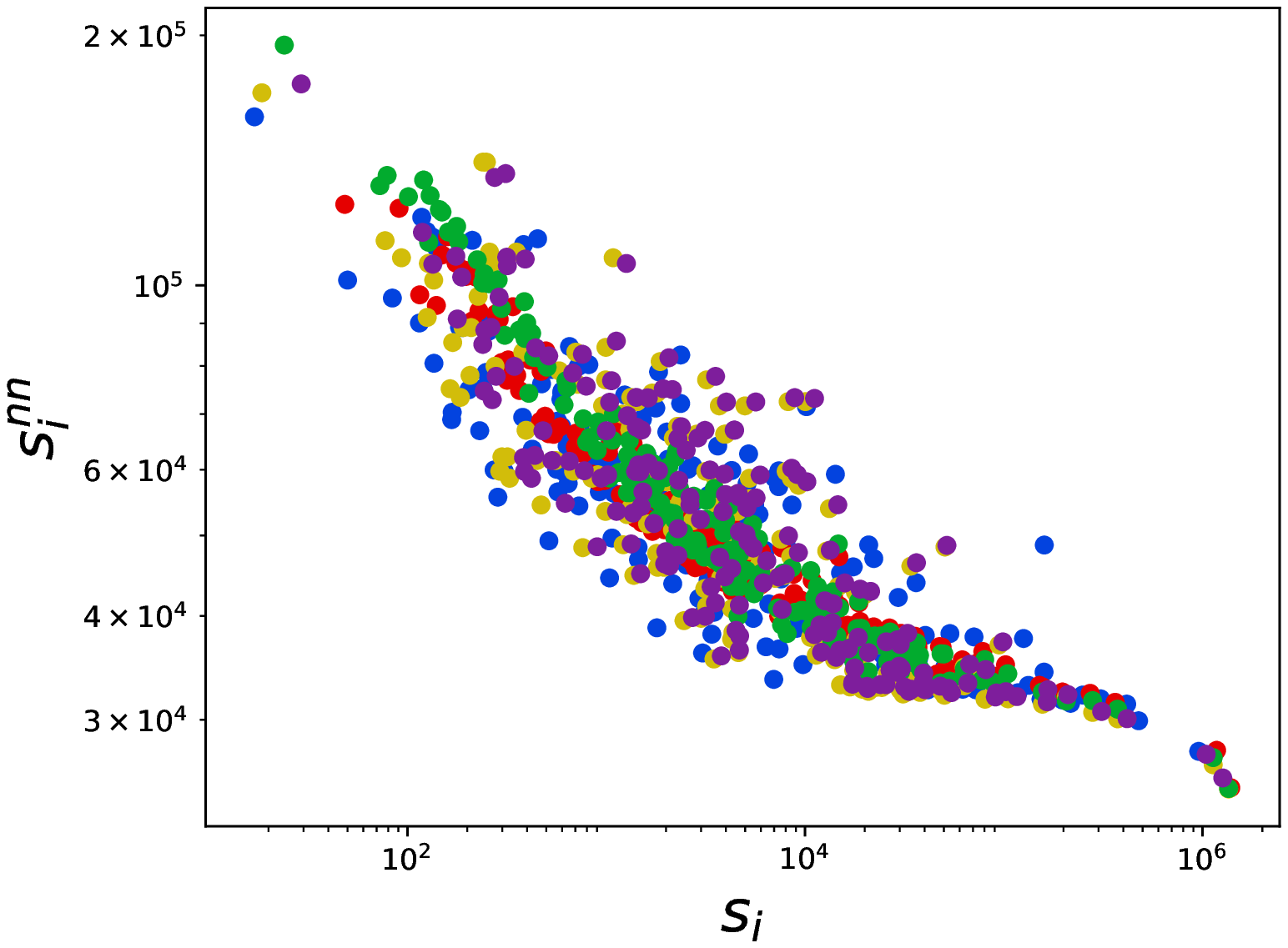}\label{fig3e}}
\quad
\subfloat[Comparison between the empirical WCC values and the ones predicted by the ME models]{\includegraphics[width=.31\textwidth]{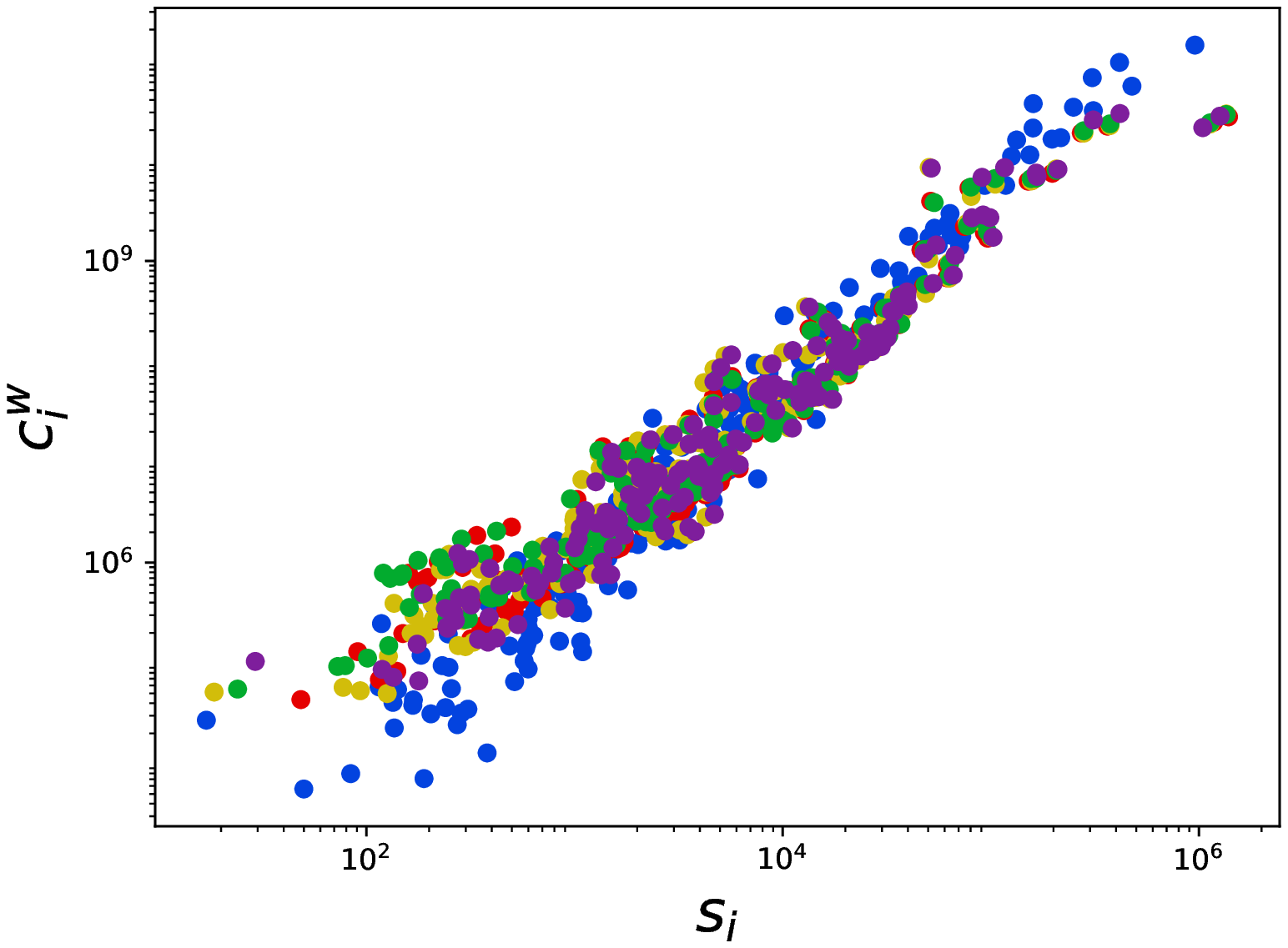}\label{fig3f}}
\caption{Performance of the econometric models versus the performance of the ME models in reproducing: (a), (d) the strengths; (b), (e) the ANNS; (c), (f) the WCC. Empirical points are indicated as $\textcolor{xkcdBlue}{\bullet}$. Econometric and ME models are indicated as follows: $\textcolor{xkcdPeriwinkle}{\bullet}$ - Poisson; $\textcolor{xkcdPink}{\bullet}$ - Negative Binomial; $\textcolor{xkcdOrange}{\bullet}$ - ZIP; $\textcolor{xkcdPeaGreen}{\bullet}$ - ZINB; $\textcolor{xkcdRed}{\bullet}$ - $H_{(1)}$; $\textcolor{xkcdPurple}{\bullet}$ - $H_{(2)}$; $\textcolor{xkcdMustardYellow}{\bullet}$ - TS; $\textcolor{xkcdKellyGreen}{\bullet}$ - TSF. Results refer to the year 2000 of the dataset curated by Gleditsch~\cite{Gleditsch2002}.}
\label{fig3}
\end{figure*}

\section{Results\label{secV}}

Let us now test and compare the performance of our two broad classes of models in reproducing the topological properties of the World Trade Web. To this aim, let us consider both the local properties, such as the degrees and the strengths, and the higher-order ones such as the average nearest neighbors degree (ANND) and the clustering coefficient (BCC), i.e.

\begin{eqnarray}
k^{nn}_i&=&\frac{\sum_{j(\neq i)} a_{ij}k_j}{k_i},\\
c_i&=&\frac{\sum_{j(\neq i,k)}\sum_{k(\neq i)}a_{ij}a_{jk}a_{ki}}{k_i(k_i-1)};
\end{eqnarray}
we will also consider their weighted counterparts, i.e. the average nearest neighbors strength (ANNS) and the weighted clustering coefficient (WCC), defined as

\begin{eqnarray}
s^{nn}_i&=&\frac{\sum_{j(\neq i)} a_{ij}s_j}{k_i},\\
c_i^w&=&\frac{\sum_{j(\neq i,k)}\sum_{k(\neq i)}w_{ij}w_{jk}w_{ki}}{k_i(k_i-1)}.
\end{eqnarray}

\begin{table*}[t!]
\centering
\begin{tabular}{C{2cm}||C{1cm}|C{1cm}|C{1cm}|C{1cm}||C{1.5cm}|C{1cm}|C{3cm}|C{1cm}}
\hline
Dataset & $H_{(1)}$ & $H_{(2)}$ & TSF & TS & Poisson & ZIP & Negative Binomial & ZINB \\
\hline
\hline
Gleditsch$_{90}$& $0.72 $& $\mathbf{0.81}$& $0.73$ & $0.79$ & $0.75$ & $0.75$ & $0.68$ & $0.70$ \\
Gleditsch$_{91}$& $0.69 $& $\mathbf{0.78}$& $0.70$ & $0.77$ & $0.73$ & $0.73$ & $0.65$ & $0.68$ \\
Gleditsch$_{92}$& $0.70 $& $\mathbf{0.79}$& $0.71$ & $0.77$ & $0.73$ & $0.73$ & $0.65$ & $0.68$ \\
Gleditsch$_{93}$& $0.70 $& $\mathbf{0.79}$& $0.71$ & $0.78$ & $0.74$ & $0.74$ & $0.66$ & $0.69$ \\
Gleditsch$_{94}$& $0.70 $& $\mathbf{0.79}$& $0.72$ & $0.78$ & $0.75$ & $0.75$ & $0.66$ & $0.69$ \\
Gleditsch$_{95}$& $0.70 $& $\mathbf{0.80}$& $0.72$ & $\mathbf{0.80}$ & $0.75$ & $0.76$ & $0.68$ & $0.69$ \\
Gleditsch$_{96}$& $0.72 $& $\mathbf{0.81}$& $0.73$ & $\mathbf{0.81}$ & $0.76$ & $0.77$ & $0.68$ & $0.69$ \\
Gleditsch$_{97}$& $0.73 $& $\mathbf{0.82}$& $0.74$ & $0.81$ & $0.77$ & $0.77$ & $0.69$ & $0.71$ \\
Gleditsch$_{98}$& $0.73 $& $\mathbf{0.82}$& $0.74$ & $0.81$ & $0.77$ & $0.77$ & $0.69$ & $0.71$ \\
Gleditsch$_{99}$& $0.73 $& $\mathbf{0.82}$& $0.74$ & $0.81$ & $0.77$ & $0.77$ & $0.69$ & $0.72$ \\
Gleditsch$_{00}$& $0.74 $& $\mathbf{0.82}$& $0.74$ & $0.81$ & $0.78$ & $0.77$ & $0.70$ & $0.72$ \\
\hline
\hline
BACI$_{07}$& $0.83 $& $\mathbf{0.88}$& $0.83$ & $0.87$ & $0.84$ & $0.84$ & $0.76$ & $0.77$ \\
BACI$_{08}$& $0.83 $& $\mathbf{0.88}$& $0.83$ & $0.87$ & $0.85$ & $0.84$ & $0.76$ & $0.77$ \\
BACI$_{09}$& $0.84 $& $\mathbf{0.88}$& $0.84$ & $0.87$ & $0.84$ & $0.84$ & $0.76$ & $0.77$ \\
BACI$_{10}$& $0.85 $& $\mathbf{0.89}$& $0.85$ & $0.88$ & $0.85$ & $0.85$ & $0.77$ & $0.78$ \\
BACI$_{11}$& $0.85 $& $\mathbf{0.89}$& $0.85$ & $0.88$ & $0.86$ & $0.85$ & $0.77$ & $0.78$ \\
BACI$_{12}$& $0.85 $& $\mathbf{0.89}$& $0.85$ & $0.88$ & $0.86$ & $0.85$ & $0.77$ & $0.78$ \\
BACI$_{13}$& $0.85 $& $\mathbf{0.90}$& $0.85$ & $0.89$ & $0.86$ & $0.86$ & $0.77$ & $0.78$ \\
BACI$_{14}$& $0.85 $& $\mathbf{0.90}$& $0.85$ & $0.89$ & $0.86$ & $0.85$ & $0.77$ & $0.78$ \\
BACI$_{15}$& $0.85 $& $\mathbf{0.90}$& $0.85$ & $0.89$ & $0.85$ & $0.84$ & $0.77$ & $0.78$ \\
BACI$_{16}$& $0.84 $& $\mathbf{0.90}$& $0.84$ & $0.89$ & $0.85$ & $0.84$ & $0.76$ & $0.78$ \\
BACI$_{17}$& $0.85 $& $\mathbf{0.90}$& $0.85$ & $0.89$ & $0.85$ & $0.85$ & $0.77$ & $0.78$ \\
\hline
\end{tabular}
\caption{Accuracy of ME and econometric models in reconstructing both the Gleditsch and the BACI dataset. The best performing models are $H_{(2)}$ and TS.}
\label{tab_accuracy_static}
\end{table*}

We will also test the accuracy of the reconstruction provided by the methods in our basket by considering properties like the true positive rate (TPR)

\begin{equation}
\langle TPR\rangle=\frac{\langle TP\rangle}{L}=\frac{\sum_{i<j}a_{ij}p_{ij}}{L}
\end{equation}
i.e. the percentage of links correctly recovered by a given reconstruction method, the specificity (SPC)

\begin{equation}
\langle SPC\rangle=\frac{\langle TN\rangle}{\binom{N}{2}-L}=\frac{\sum_{i<j}(1-a_{ij})(1-p_{ij})}{\binom{N}{2}-L}
\end{equation}
i.e. the percentage of zeros correctly recovered by a given reconstruction method, the positive predictive value (PPV) 

\begin{equation}
\langle PPV\rangle=\frac{\langle TP\rangle}{\langle L\rangle}=\frac{\sum_{i<j}a_{ij}p_{ij}}{\langle L\rangle}
\end{equation}
i.e. the percentage of links correctly recovered by a given reconstruction method with respect to the total number of predicted links and the accuracy (ACC)

\begin{equation}
\langle ACC\rangle=\frac{\langle TP\rangle+\langle TN\rangle}{\binom{N}{2}}
\end{equation}
measuring the overall performance of a given reconstruction method in correctly placing both links and zeros.

Fig.~\ref{fig2} sums up the comparisons carried out between the econometric models and the ME ones. The comparison between the empirical cumulative density function (CDF) of the degrees and the ones output by the econometric models reveals the latter ones to be able to predict an overall similar functional form (see Fig.~\ref{fig2}\subref{fig2a} and Fig.~\ref{fig2}\subref{fig2d}); still, the prediction obtained by any of the ME models is much closer to the empirical trend. More quantitatively, one can implement the Kolmogorov-Smirnov (KS) test to check the goodness of any of the models considered in the present work to replicate the empirical degrees: while any of the ME models provides estimates of the degrees that are compatible with the empirical CDF (at the significance level of $5\%$), only the ZIP model predicts degrees that are compatible with the empirical ones: in fact, the p-values of the ME models read $p_{(1)}\simeq 0.06$, $p_{(2)}\simeq 0.99$, $p_\text{TS}\simeq 0.99$, $p_\text{TSF}\simeq 0.32$ while the p-values of the econometric models read $p_\text{Pois}\simeq 0.001$, $p_\text{NB}\simeq 0.0008$, $p_\text{ZIP}\simeq 0.63$, $p_\text{ZINB}\simeq 0.001$.

Coming to the higher-order properties, it is evident that the majority of the econometric models fails to overlap with the empirical cloud of points (see Fig.~\ref{fig2}\subref{fig2b}): the one providing the best prediction is the ZIP model, whose performance represents quite an improvement with respect to the one provided by the `plain' Poisson model. While this is quite evident for what concerns the prediction of the ANND values, the performances of the ZIP and of the `plain' Poisson model become less different when tested on the BCC values. On the contrary, the performance of the ZINB model closely resembles that of the negative binomial one when tested both on the ANND and on the BCC. As for the local properties, the KS test reveals that the only model outputting predictions compatible with the empirical values (at the significance level of $5\%$) is the ZIP one: in fact, $p_\text{ZIP}^\text{ANND}\simeq 0.38$, $p_\text{ZIP}^\text{BCC}\simeq 0.08$).

For what concerns ME models, the ones performing best are those constraining the degrees, i.e. the model induced by $H_{(2)}$ and its two-step counterpart, whose topological estimation step is carried out by employing $p_{ij}^\text{UBCM}$. The evidence that their performances in reproducing the purely binary structure of a network are very similar lets us suspect that $p_{ij}^{(2)}\simeq p_{ij}^\text{UBCM}$ and conclude that the purely econometric information encoded into $p_{ij}^{(2)}$ does not add much to what is already conveyed by the purely topological one. On the other hand, ME models not constraining the degrees provide predictions differing from the empirical trends to quite a large extent. As the KS test reveals, the only ME model outputting predictions that are not compatible with the empirical values (at the significance level of $5\%$) is the one induced by $H_{(1)}$.

The overall accuracy of our models in reproducing a network topology can be proxied by the index $\Delta_L=|\langle L\rangle-L|/L$ amounting at $\Delta^\text{Pois}_L\gtrsim\Delta^\text{NB}_L=\Delta^\text{ZINB}_L\simeq 6\%$ while $\Delta^\text{ZIP}_L\simeq 0.5\%$ and $\Delta^\text{ME}_L=0$ for each ME model. This is confirmed by our analysis of single link statistics: in fact, $\langle ACC\rangle_{(2)}\simeq 0.83$ attains the largest value, followed by $\langle ACC\rangle_\text{TS}\simeq 0.81$ and $\langle ACC\rangle_\text{ZIP}\simeq 0.77$. Remarkably, $\langle PPV\rangle_{(2)}\simeq 0.86$ attains the largest value, indicating that the ME model induced by $H_{(2)}$ is the one placing links best among all the models in our basket.\\

Let us now consider the weighted properties (see  Fig.~\ref{fig3}). Overall, the distribution of the strengths is reproduced quite well by all models considered here, although no one explicitly constrains them. This seems to indicate that the purely econometric information `feeded' into our models indeed plays a role - which, however, is limited to ensure that the intensive margins (and the related properties, as we will see) are accurately predicted. The larger explanatory power of econometric models becomes now evident: all of them output predictions that are compatible with the empirical values. Although the same result holds true for ME models, the latter ones are outperformed by purely econometric models - the best performing ones in predicting the strengths being the Poisson-like ones.

Coming to the higher-order properties, let us notice that the best performing econometric models in reproducing the ANNS values are the ZIP and the `plain' Poisson ones whose performances differ less than in the ANND case - although the KS test lets the ZIP model win. On the other hand, the ZINB and the negative binomial models (whose performances are, again, very similar) completely fail in capturing the empirical values. All predictions from ME models overlap with the empirical ANNS values: as the KS test reveals, the only ME model outputting predictions that are not compatible with the empirical values (at the significance level of $1\%$) is the one induced by $H_{(1)}$. For what concerns the values of the WCC, both the econometric and the ME models perform quite satisfactorily in capturing its rising trend. However, the KS test reveals that only the econometric models and the TS model output predictions compatible with the WCC empirical values (at the significance level of $1\%$).

To proxy the accuracy of our models in reproducing the weighted network structure we have considered the index $\Delta_W=|\langle W\rangle-W|/W$ amounting at $\Delta^\text{ZINB}_W\simeq 95\%$, $\Delta^\text{NB}_W=60\%$, $\Delta^\text{ZIP}_W\simeq 0.3\%$ and $\Delta^\text{Pois}_W\simeq 0$ while $\Delta^\text{ME}_W=0$ for the ME models constraining the total weight and $\Delta^\text{TS}_W\gtrsim\Delta^\text{TSF}_W\simeq 0.2\%$ for the ME two-step ones.

\begin{figure*}
\centering
\includegraphics[width=\textwidth]{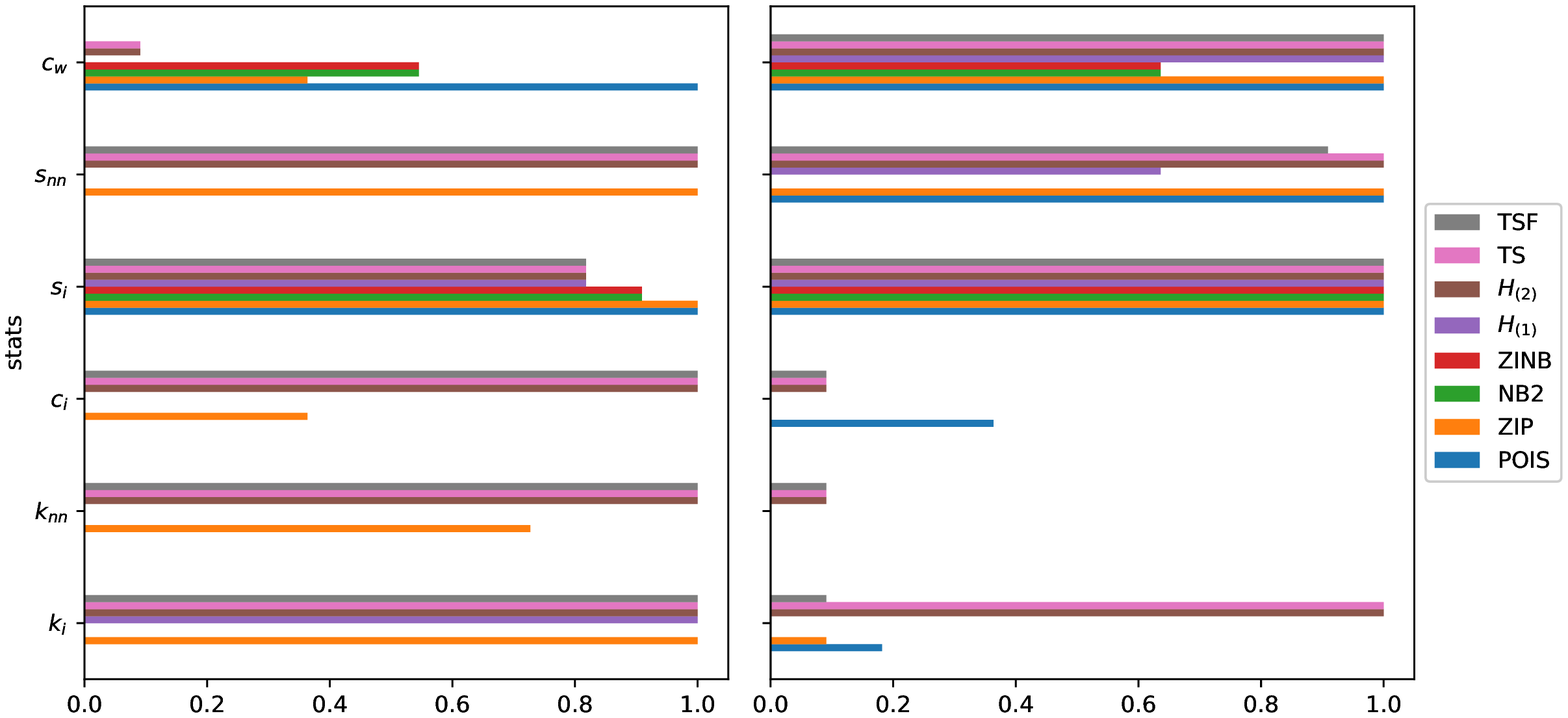}
\caption{Percentage of times the empirical value of a given network statistics is compatible with its ensemble distribution, according to a KS test at the significance level of $5\%$ (left, for the Gleditsch dataset; right, for the BACI dataset): a value of 1 means that the given network statistic is compatible with the model-induced ensemble distribution across all years of the considered dataset. Generally speaking, ME models are the ones for which compatibility is largest - although econometric models are characterized by a compatibility which is large as well whenever weighted measures are considered.}
\label{netstats_ks_frequencies}
\end{figure*}

In order to understand if the conclusions above can be generalized, let us calculate the accuracy of all models in our basket, for all years constituting our two datasets. The results, summed up in Tab.~\ref{tab_accuracy_static}, confirm that model $H_{(2)}$ systematically outperforms all competing models. As an additional test, we have calculated the percentage of times the empirical values of our network statistics are compatible with their ensemble distributions, via KS tests at the significance level of $5\%$: the results, shown in Fig.~\ref{netstats_ks_frequencies}, point out that ME models are the ones for which compatibility is largest.\\

Let us now ask ourselves if a criterion exist to carry out a principled comparison of the performance of the models considered in the present work. The answer is positive and lays in the adoption of the popular \textit{Akaike Information Criterion} and \textit{Bayesian Information Criterion}, respectively defined as

\begin{equation}
\text{AIC}=2K-2\mathcal{L}
\end{equation}

and
\begin{equation}
\text{BIC}=K\ln n-2\mathcal{L}
\end{equation}
where $\mathcal{L}$ is the log-likelihood of the tested model evaluated in its maximum, $K$ is the number of parameters characterizing the model itself and $n$ is the cardinality of the set of observations - estimated as $\frac{N(N-1)}{2}$ for undirected network data. Model selection based on these criteria prescribes to rank models according to (either) their AIC or BIC value and choose the one minimizing it. Tab.~\ref{tab1} shows both the AIC and the BIC values for all the models considered here.

Quite surprisingly, the negative binomial model is the favoured one among the econometric models, followed by its zero-inflated version; however, its bad performance in reproducing the empirical trends makes the choice of including it among the most suitable models for modelling trade data highly questionable. On the other hand, the ZIP model performs much better in reconstructing the trends of both local and higher-order properties although being much less parsimonious than both versions of the negative binomial model. Apparently, then, the question about which model to prefer - i.e. the favoured one by information criteria or the best performing one in reproducing trends? - cannot be properly answered by just considering purely econometric models. On the other hand, such a question can be unambiguously answered as soon as one switches to the class of maximum-entropy models: now, both the AIC- and the BIC-based rankings favour the model described by the Hamiltonian $H_{(2)}$ - the one encoding the information about the degree sequence and the total weight, plus admitting a tunable function of the weights - i.e. precisely the most accurate in replicating many (if not all) empirical trends.

For the sake of comparison, we have included into the basket of maximum-entropy models the Undirected Enhanced Configuration Model (UECM), i.e. the model performing best in presence of complete information about the constraints - degrees and strengths, in the specific case - of a given networked system: as evident from the table, it is disfavoured with respect to the model described by the Hamiltonian $H_{(2)}$, an evidence signalling that while the information encoded into the degrees is essential (i.e. the latter ones must be explicitly constrained), the one carried by the strengths appears to be `less fundamental' since providing a good approximation of them is enough to obtain an overall good reconstruction.

\begin{table*}[t!]
\centering
\begin{tabular}{C{3.5cm}|C{3cm}|C{2cm}|C{2cm}|C{2cm}}
\hline
Model & $\Delta_L$ & $\Delta_W$ & AIC & BIC \\
\hline
\hline
Poisson & $\simeq 0.07$ & $0^*$ & $\simeq 5088080.1$ & $\simeq 5088105.1$ \\
ZI Poisson & $\simeq 4.5\cdot 10^{-3}$ & $\simeq 3.3 \cdot 10^{-3}$ & $\simeq 5052767.3$ & $\simeq 5052800.6$ \\
{\bf Negative Binomial} & $\simeq 0.06$ & $\simeq0.59$ & $\simeq \mathbf{175693.2}$ & $\simeq \mathbf{175726.6}$ \\
ZI Negative Binomial & $\simeq 0.06$ & $\simeq 0.95$ & $\simeq 176071.4$ & $\simeq 176113.1$ \\
\hline
\hline
ME model $H_{(1)}$ & $0^*$ & $0^*$ & $\simeq 182909.5$ & $\simeq 182951.2$ \\
{\bf ME model} $H_{(2)}$ & $0^*$ & $0^*$ & $\simeq \mathbf{174050.0} $ & $\simeq \mathbf{175550.4}$ \\
TS & $0^*$ & $\simeq6.8 \cdot 10^{-3}$ & $\simeq 176129.1$ & $\simeq 177629.5$ \\
TSF & $0^*$ & $\simeq2.2 \cdot 10^{-3}$  & $\simeq 185940.2$ & $\simeq 185981.8$ \\
\hline
\hline
UECM & $0^*$ & $0^*$ & $\simeq 186602.8$ & $\simeq 189536.8$ \\
\hline
\end{tabular}
\caption{Selection of the econometric and the maximum-entropy models, according to the AIC and BIC values. While the negative binomial model is the favoured one among the econometric models, it performs badly in reproducing both local and higher-order topological properties; on the other hand, the ZIP model reproduces the higher-order statistics quite accurately but is disfavoured by both AIC and BIC. The solution to this dilemma comes from considering a different class of reconstruction models, i.e. the maximum-entropy ones: beside being the favoured one by information criteria, the $H_{(2)}$ model achieves a very good reconstruction of both binary and weighted network properties (since $\Delta L=|\langle L\rangle-L|/L$ and $\Delta W=|\langle W\rangle-W|/W$, the symbol $0^*$ indicates that the model exactly reproduces the corresponding constraint). Results refer to the year 2000 of the dataset curated by Gleditsch~\cite{Gleditsch2002}.}
\label{tab1}
\end{table*}

In order to understand if the conclusions above can be generalized, let us calculate the Akaike weights for the models in our basket $\mathcal{M}$. The Akaike weight for the $i$-th model is defined as

\begin{equation}
w_i=\frac{e{^{-\Delta_i/2}}}{\sum_m e{^{-\Delta_m/2}}},
\end{equation}
with $\Delta_i=\text{AIC}_i-\min\{\text{AIC}_m\}_{m\in\mathcal{M}}$. Results on the dataset curated by Gleditsch show that the negative binomial and the $H_{(2)}$ models `compete', in the sense that $H_{(2)}$ performs best (i.e. $w_{H_{(2)}}\simeq 1$) in the (bunches of) years 1990-1993 and 1997-2000 while the negative binomial model performs best (i.e. $w_{NB}\simeq 1$) in the (bunch of) years 1994-1996. For what concerns the BACI dataset, instead, the competing models are three: in fact, while $H_{(2)}$ performs best in the (bunches of) years 2007, 2009 and 2015-2017, the negative binomial outperforms the others in the (bunches of) years 2008 and 2010-2014; however, the ZINB model has a positive, non-negligible Akaike weight in the years 2008, 2012 and 2014, hence performing as well as the negative binomial one.\\

Let us now consider a couple of additional exercises, carried out on both datasets considered in the present work.

The first one concerns link prediction and was carried out by following the reference~\cite{Ashraf2019}. Specifically, we have approached link prediction from a temporal perspective, inspecting the accuracy achieved by our reconstruction models at time $t+1$ given the knowledge about the network topology (for the maximum-entropy models) and of the other exogenous variables (for the purely econometric models) at time $t$. In other words, we opted for a one-lagged link prediction, calculating the log-likelihood

\begin{equation}
\mathcal{L}_{1l}=\ln\left[\prod_{i<j}\left(p_{ij}^{(t)}\right)^{a_{ij}^{(t+1)}}\left(1-p_{ij}^{(t)}\right)^{1-a_{ij}^{(t+1)}}\right]
\end{equation}
for each statistical model in our basket, the coefficients $\left\{p_{ij}^{(t)}\right\}_{i,j=1}^N$ being the probabilities output by any model at time $t$ and the coefficients $\left\{a_{ij}^{(t+1)}\right\}_{i,j=1}^N$ being the entries of the adjacency matrix at time $t+1$. When carried out on the pairs of years 1993-1994, 1994-1995, 1995-1996, 1996-1997, 1997-1998, 1998-1999 and 1999-2000 of the dataset curated by Gleditsch and on the pairs of years 2008-2009, 2009-2010, 2010-2011, 2013-2014, 2014-2015,2015-2016 and 2016-2017 of the BACI dataset, the exercise above shows $H_{(2)}$ to outperform not only the entire class of econometric models but also the purely binary, maximum-entropy ones - as confirmed by the Akaike weights induced by the log-likelihood above.

To provide a more refined picture of the performance of the models in our basket in providing one-lagged predictions, we have also calculated their one-lagged accuracy, defined as

\begin{equation}
\langle ACC\rangle_{1l}=\frac{\langle TP\rangle_{1l}+\langle TN\rangle_{1l}}{\binom{N}{2}}
\end{equation}
with $\langle TP\rangle_{1l}=\sum_{i<j}a_{ij}^{(t+1)}p_{ij}^{(t)}$ and $\langle TN\rangle_{1l}=\sum_{i<j}\left(1-a_{ij}^{(t+1)}\right)\left(1-p_{ij}^{(t)}\right)$. The results are reported in Tab.~\ref{1laggedaccuracy} and confirm what has been previously said: $H_{(2)}$ is the one performing best.

\begin{table*}[t!]
\centering
\begin{tabular}{C{2cm}||C{1cm}|C{1cm}|C{1cm}|C{1cm}||C{1.5cm}|C{1cm}|C{3cm}|C{1cm}}
\hline
Dataset & $H_{(1)}$ & $H_{(2)}$ & TSF & TS & Poisson & ZIP & Negative Binomial & ZINB \\
\hline
\hline
Gleditsch$_{94}$& $0.70 $& $\mathbf{0.79}$& $0.72$ & $0.77$ & $0.75$ & $0.74$ & $0.66$ & $0.69$ \\
Gleditsch$_{95}$& $0.71 $& $\mathbf{0.80}$& $0.72$ & $0.78$ & $0.75$ & $0.75$ & $0.66$ & $0.69$ \\
Gleditsch$_{96}$& $0.72 $& $\mathbf{0.80}$& $0.73$ & $\mathbf{0.80}$ & $0.76$ & $0.76$ & $0.67$ & $0.70$ \\
Gleditsch$_{97}$& $0.73 $& $\mathbf{0.82}$& $0.74$ & $0.80$ & $0.77$ & $0.77$ & $0.67$ & $0.70$ \\
Gleditsch$_{98}$& $0.73 $& $\mathbf{0.82}$& $0.74$ & $0.81$ & $0.77$ & $0.77$ & $0.69$ & $0.71$ \\
Gleditsch$_{99}$& $0.73 $& $\mathbf{0.82}$& $0.74$ & $0.81$ & $0.77$ & $0.77$ & $0.69$ & $0.71$ \\
Gleditsch$_{00}$& $0.73 $& $\mathbf{0.82}$& $0.74$ & $0.81$ & $0.77$ & $0.77$ & $0.69$ & $0.71$ \\
\hline
\hline
BACI$_{09}$& $0.84 $& $\mathbf{0.88}$& $0.84$ & $0.87$ & $0.85$ & $0.84$ & $0.76$ & $0.77$ \\
BACI$_{10}$& $0.84 $& $\mathbf{0.88}$& $0.84$ & $0.87$ & $0.85$ & $0.84$ & $0.76$ & $0.77$ \\
BACI$_{11}$& $0.85 $& $\mathbf{0.89}$& $0.85$ & $0.88$ & $0.86$ & $0.85$ & $0.77$ & $0.78$ \\
BACI$_{14}$& $0.85 $& $\mathbf{0.89}$& $0.85$ & $0.88$ & $0.86$ & $0.85$ & $0.77$ & $0.78$ \\
BACI$_{15}$& $0.85 $& $\mathbf{0.89}$& $0.85$ & $0.88$ & $0.86$ & $0.85$ & $0.77$ & $0.78$ \\
BACI$_{16}$& $0.84 $& $\mathbf{0.88}$& $0.84$ & $0.88$ & $0.85$ & $0.84$ & $0.77$ & $0.78$ \\
BACI$_{17}$& $0.85 $& $\mathbf{0.89}$& $0.85$ & $0.88$ & $0.85$ & $0.84$ & $0.77$ & $0.78$ \\
\hline
\end{tabular}
\caption{One-lagged accuracy, quantifying the performance of ME and econometric models in providing one-lagged predictions on both the Gleditsch and the BACI datasets. $H_{(2)}$ and TS are systematically the best performing models.}
\label{1laggedaccuracy}
\end{table*}

As a second exercise, we have tested the accuracy of our models in estimating link-specific weights - hence, carrying out what we have called a `weight prediction' exercise. To this aim, we have considered each expected weight and calculated the confidence interval enclosing the $95\%$ of total probability around it. On the practical side, we have sampled 1000 configurations from the ensemble induced by each model in our basket and calculated the (ensemble-induced) 2.5 and 97.5 percentiles for each specific weight; then, we have calculated the percentage of empirical weights `falling' within the corresponding CIs - now treated as error bars `accompanying' the point-estimate of each weight. The results of this exercise are reported in in Tab.~\ref{weighted_prediction_table}: as it can be appreciated, maximum-entropy models compete with both the negative binomial and the ZINB ones - although the latter (slightly) outperform the former.

\section{Discussion\label{secVI}}

In the present work, we have compared the performance of two broad classes of statistical models, i.e. the ones rooted into economic theory and the ones rooted into statistical physics (in particular, the ones derived from the maximum-entropy principle) in reconstructing both the binary and the weighted network properties of an economic system such as the WTW.

Although the case study is the same, the two classes of models `reflect' the languages of two disciplines that are still deeply different: while econometricians have traditionally focused on bilateral trade volumes between countries - emphasizing the role played by common borders, language, religion, the presence of regional trade agreements, etc. on trade relationships - network scientists have, instead, paid more attention to the structural and dynamical aspects of network formation, emphasizing the role played by purely structural information in determining the topology itself. The 2008 global financial crisis has dramatically clarified that bilateral trade relationships can explain only a small fraction of the impact that an economic shock, originating in a given country, can have on another country which is not a direct trade partner, urging researchers in economics to adopt a network-aware perspective. This, in turn, has motivated us to carry out a methodological comparison on real-world cases, with the aim of clarifying pros and cons of both approaches.

Researchers in economics have dealt with the issue of reconstructing network topology by approaching the simpler problem of reproducing the number of missing connections - or, equivalently, the link density. For instance, as we see from the year 2000 snapshot of the dataset curated by Gleditsch~\cite{Gleditsch2002}, although the error of the Poisson model in reproducing $L$ is overall small (amounting at $\Delta^\text{Pois}_L\simeq 7\%$), it can be further reduced by adopting the zero-inflated version of it. On the contrary, inflating zeros does not improve the performance of the negative binomial model in reproducing the link density since it already underestimates $L$. The ability of a model in reproducing a global quantity such as the link density proxies its ability in providing a good estimation of local as well as higher-order topological properties (i.e. the degrees, the ANND and the BCC): from this point of view, the zero-inflated Poisson model is the one performing best among the econometric models. However, it is largely disfavoured by information criteria such as AIC and BIC, a result suggesting that it may be not parsimonious enough.

\begin{table*}[t!]
\centering
\begin{tabular}{C{2cm}||C{1cm}|C{1cm}|C{1cm}|C{1cm}||C{1.5cm}|C{1cm}|C{3cm}|C{1cm}}
\hline
Dataset & $H_{(1)}$ & $H_{(2)}$ & TSF & TS & Poisson & ZIP & Negative Binomial & ZINB \\
\hline
\hline
Gleditsch$_{90}$& $0.96 $& $0.96$& $0.94$ & $0.96$ & $0.62$ & $0.67$ & $\mathbf{0.98}$ & $0.97$ \\
Gleditsch$_{91}$& $0.96 $& $0.96$& $0.94$ & $0.96$ & $0.63$ & $0.71$ & $\mathbf{0.98}$ & $0.97$ \\
Gleditsch$_{92}$& $0.96 $& $0.96$& $0.94$ & $0.96$ & $0.63$ & $0.70$ & $\mathbf{0.98}$ & $0.97$ \\
Gleditsch$_{93}$& $0.96 $& $0.96$& $0.94$ & $0.96$ & $0.64$ & $0.69$ & $\mathbf{0.98}$ & $0.97$ \\
Gleditsch$_{94}$& $0.96 $& $0.96$& $0.94$ & $0.96$ & $0.64$ & $0.67$ & $\mathbf{0.98}$ & $0.97$ \\
Gleditsch$_{95}$& $0.96 $& $0.96$& $0.94$ & $0.96$ & $0.61$ & $0.68$ & $\mathbf{0.98}$ & $0.97$ \\
Gleditsch$_{96}$& $0.96 $& $0.96$& $0.94$ & $0.96$ & $0.60$ & $0.65$ & $\mathbf{0.98}$ & $0.97$ \\
Gleditsch$_{97}$& $0.96 $& $0.96$& $0.94$ & $0.96$ & $0.60$ & $0.63$ & $\mathbf{0.98}$ & $0.97$ \\
Gleditsch$_{98}$& $0.96 $& $0.96$& $0.94$ & $0.96$ & $0.61$ & $0.63$ & $\mathbf{0.98}$ & $0.97$ \\
Gleditsch$_{99}$& $0.96 $& $0.96$& $0.94$ & $0.96$ & $0.61$ & $0.62$ & $\mathbf{0.98}$ & $0.97$ \\
Gleditsch$_{00}$& $0.96 $& $0.96$& $0.95$ & $0.96$ & $0.60$ & $0.63$ & $\mathbf{0.97}$ & $\mathbf{0.97}$ \\
\hline
\hline
BACI$_{07}$& $0.93 $& $0.95$& $0.92$ & $0.95$ & $0.43$ & $0.43$ & $\mathbf{0.97}$ & $\mathbf{0.97}$ \\
BACI$_{08}$& $0.92 $& $0.94$& $0.91$ & $0.94$ & $0.40$ & $0.40$ & $\mathbf{0.97}$ & $\mathbf{0.97}$ \\
BACI$_{09}$& $0.93 $& $0.95$& $0.92$ & $0.95$ & $0.43$ & $0.43$ & $\mathbf{0.97}$ & $\mathbf{0.97}$ \\
BACI$_{10}$& $0.92 $& $0.94$& $0.91$ & $0.95$ & $0.41$ & $0.41$ & $\mathbf{0.97}$ & $\mathbf{0.97}$ \\
BACI$_{11}$& $0.92 $& $0.94$& $0.90$ & $0.94$ & $0.38$ & $0.39$ & $\mathbf{0.97}$ & $\mathbf{0.97}$ \\
BACI$_{12}$& $0.91 $& $0.94$& $0.90$ & $0.94$ & $0.38$ & $0.38$ & $\mathbf{0.97}$ & $\mathbf{0.97}$ \\
BACI$_{13}$& $0.91 $& $0.93$& $0.90$ & $0.93$ & $0.37$ & $0.38$ & $\mathbf{0.97}$ & $\mathbf{0.97}$ \\
BACI$_{14}$& $0.91 $& $0.93$& $0.90$ & $0.94$ & $0.38$ & $0.39$ & $\mathbf{0.97}$ & $\mathbf{0.97}$ \\
BACI$_{15}$& $0.91 $& $0.94$& $0.91$ & $0.94$ & $0.41$ & $0.41$ & $\mathbf{0.97}$ & $\mathbf{0.97}$ \\
BACI$_{16}$& $0.92 $& $0.94$& $0.91$ & $0.94$ & $0.40$ & $0.41$ & $\mathbf{0.97}$ & $\mathbf{0.97}$ \\
BACI$_{17}$& $0.91 $& $0.94$& $0.91$ & $0.94$ & $0.39$ & $0.39$ & $\mathbf{0.97}$ & $\mathbf{0.97}$ \\
\hline
\end{tabular}
\caption{Performance of ME and econometric models in providing accurate predictions of weights, on both the Gleditsch and the BACI datasets, quantified by calculating the percentage of empirical weights falling within the confidence interval enclosing the $95\%$ probability around the corresponding expected value. Our maximum-entropy models compete with both the negative binomial and the ZINB ones - although the latter (slightly) outperform the former.}
\label{weighted_prediction_table}
\end{table*}

\begin{figure*}[t!]
\subfloat[Comparison between the empirical degree distribution and the ones predicted by econometric and ME models]{\includegraphics[width=.31\textwidth]{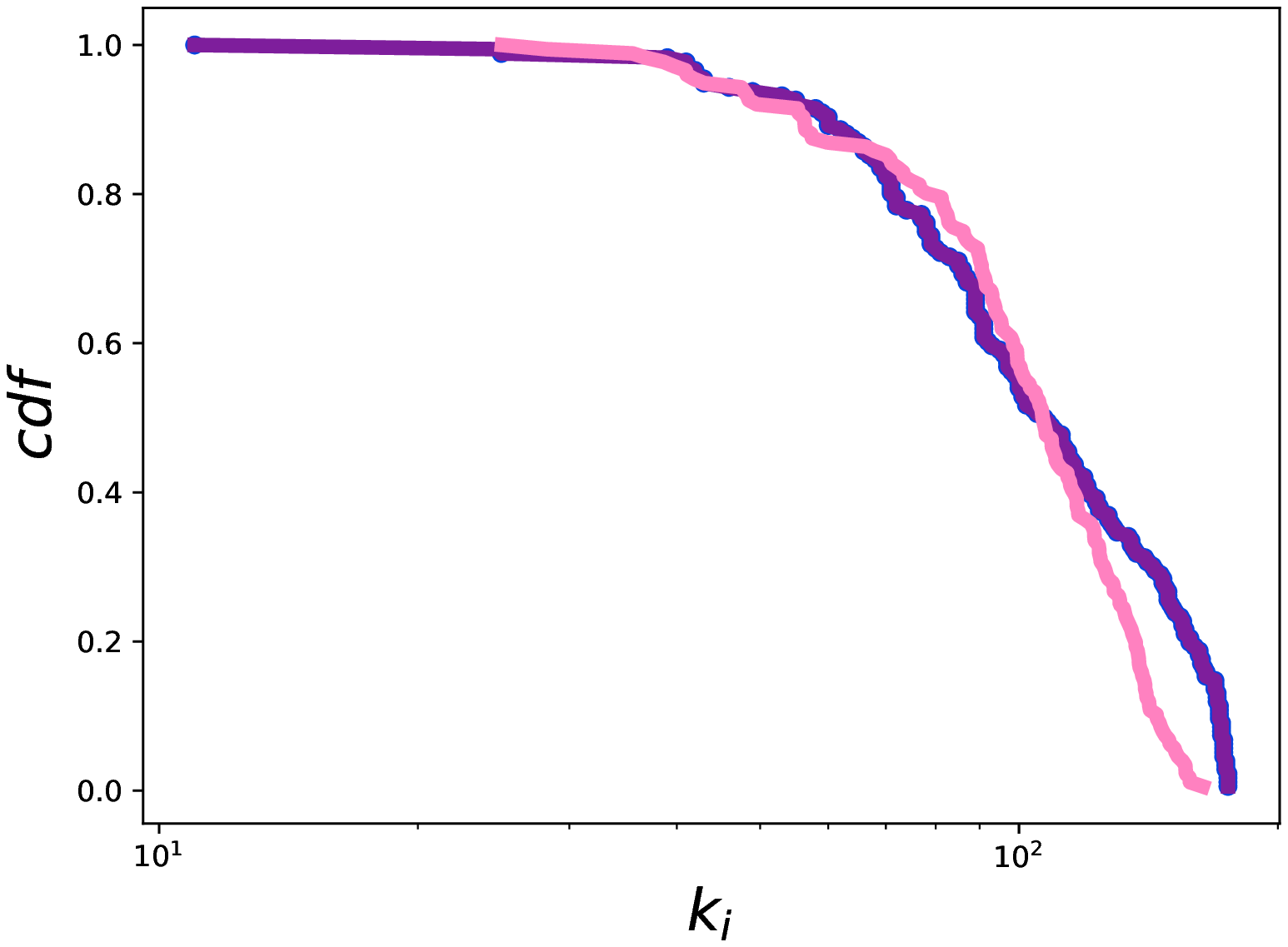}\label{fig5a}}
\quad
\subfloat[Comparison between the empirical ANND values and the ones predicted by econometric and ME models]{\includegraphics[width=.31\textwidth]{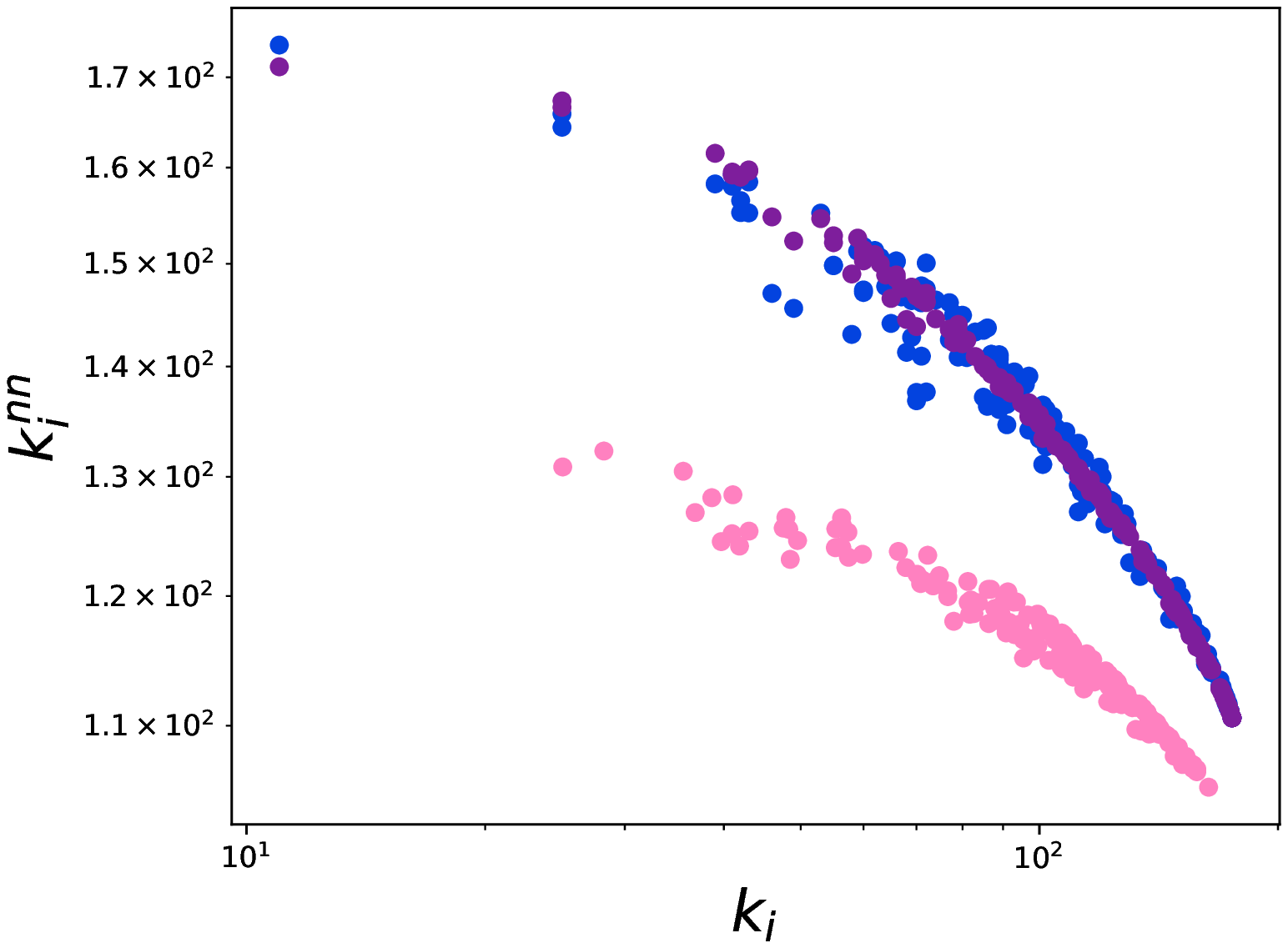}\label{fig5b}}
\quad
\subfloat[Comparison between the empirical BCC values and the ones predicted by econometric and ME models]{\includegraphics[width=.31\textwidth]{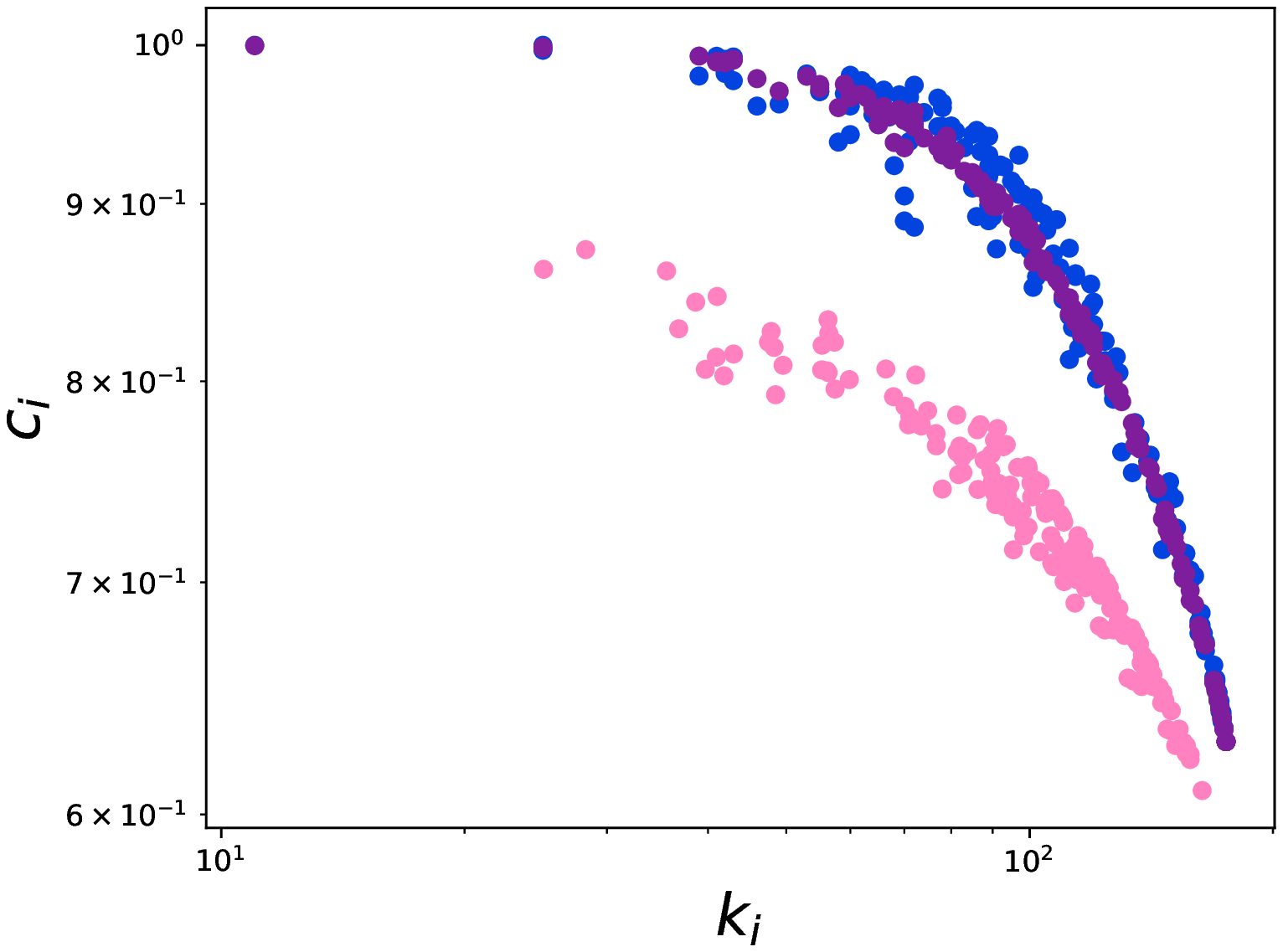}\label{fig5c}}\\

\subfloat[Comparison between the empirical strength distribution and the ones predicted by econometric and ME models]{\includegraphics[width=.31\textwidth]{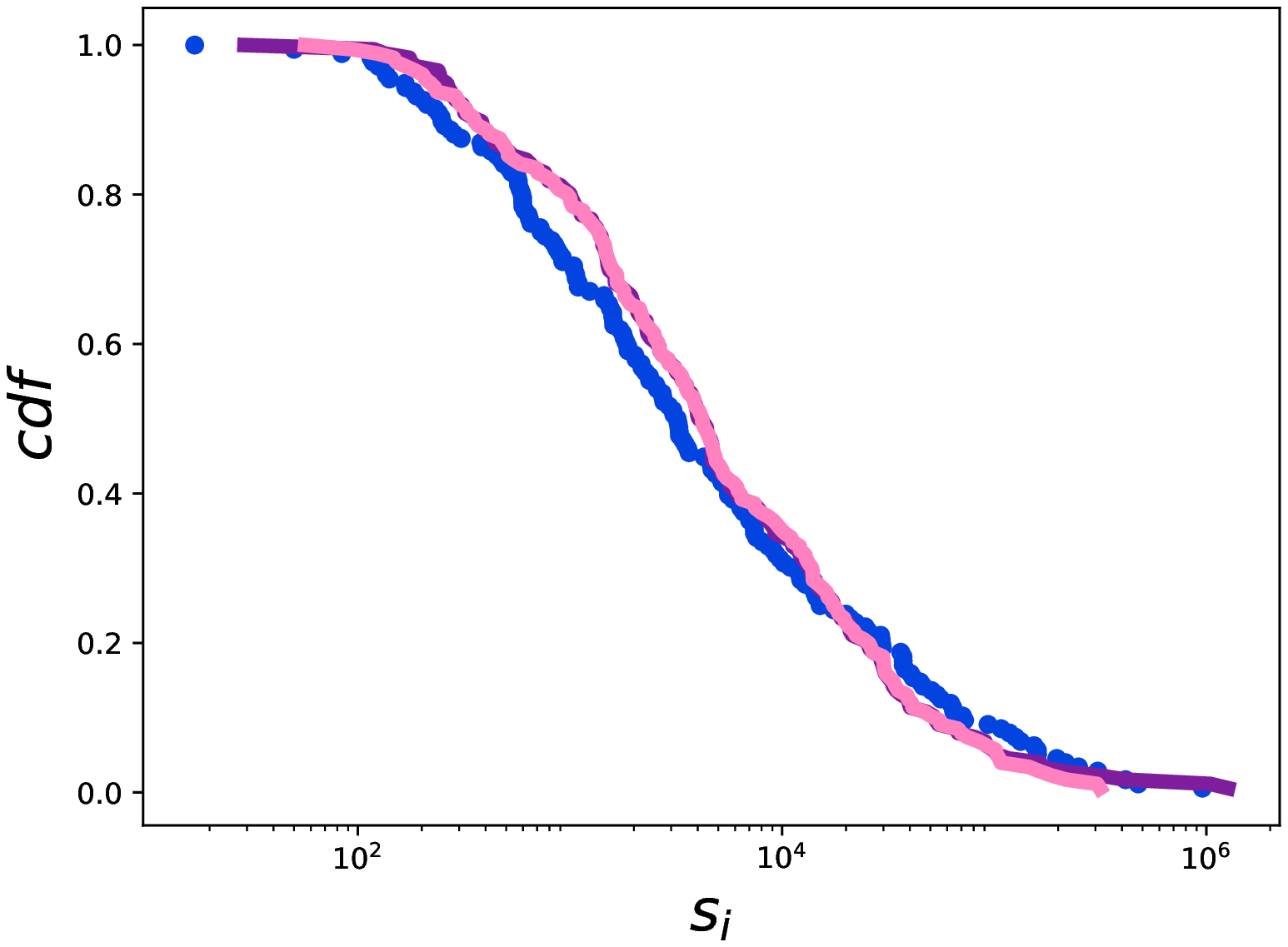}\label{fig5d}}
\quad
\subfloat[Comparison between the empirical ANNS values and the ones predicted by econometric and ME models]{\includegraphics[width=.31\textwidth]{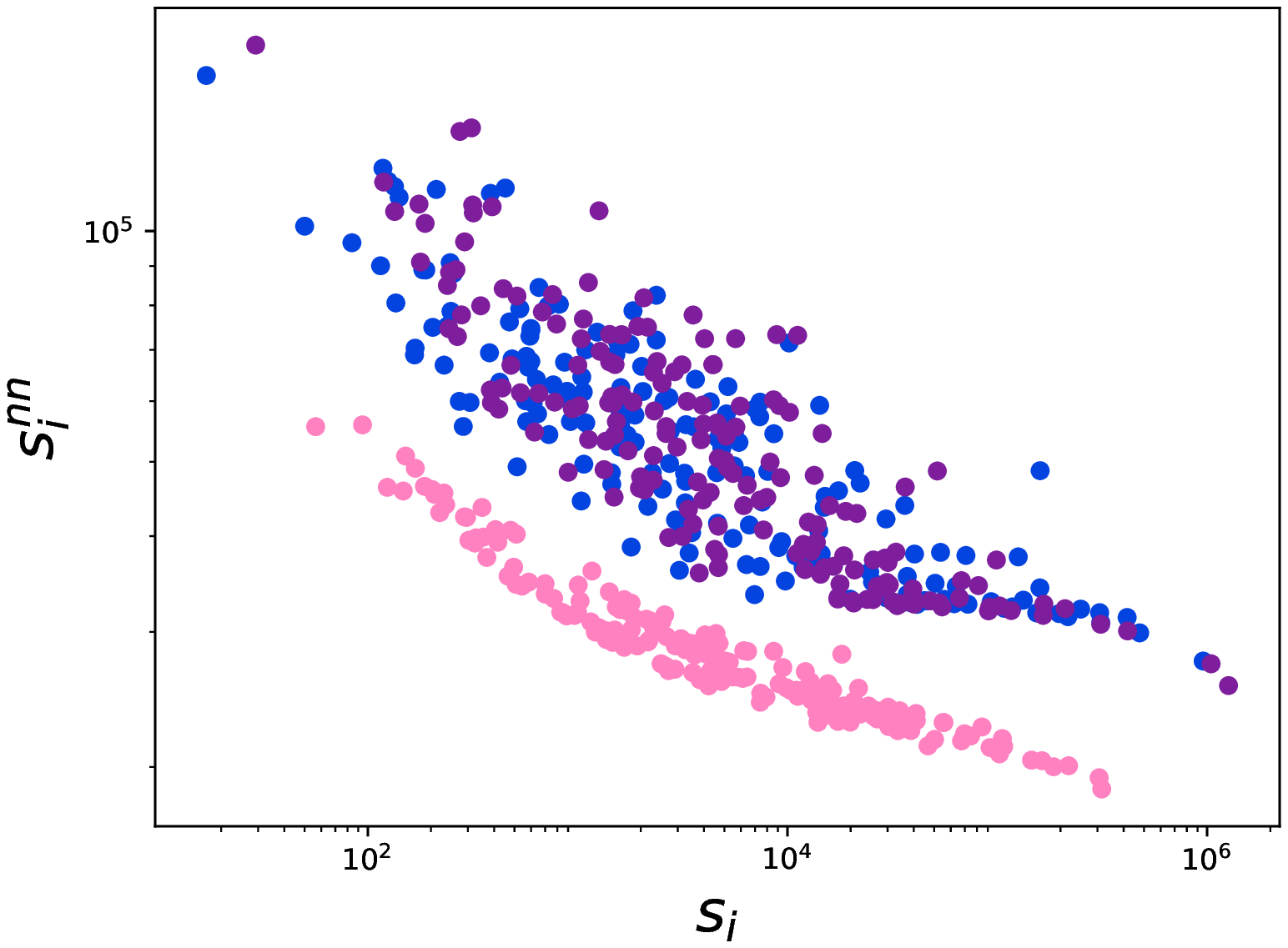}\label{fig5e}}
\quad
\subfloat[Comparison between the empirical WCC values and the ones predicted by econometric and ME models]{\includegraphics[width=.31\textwidth]{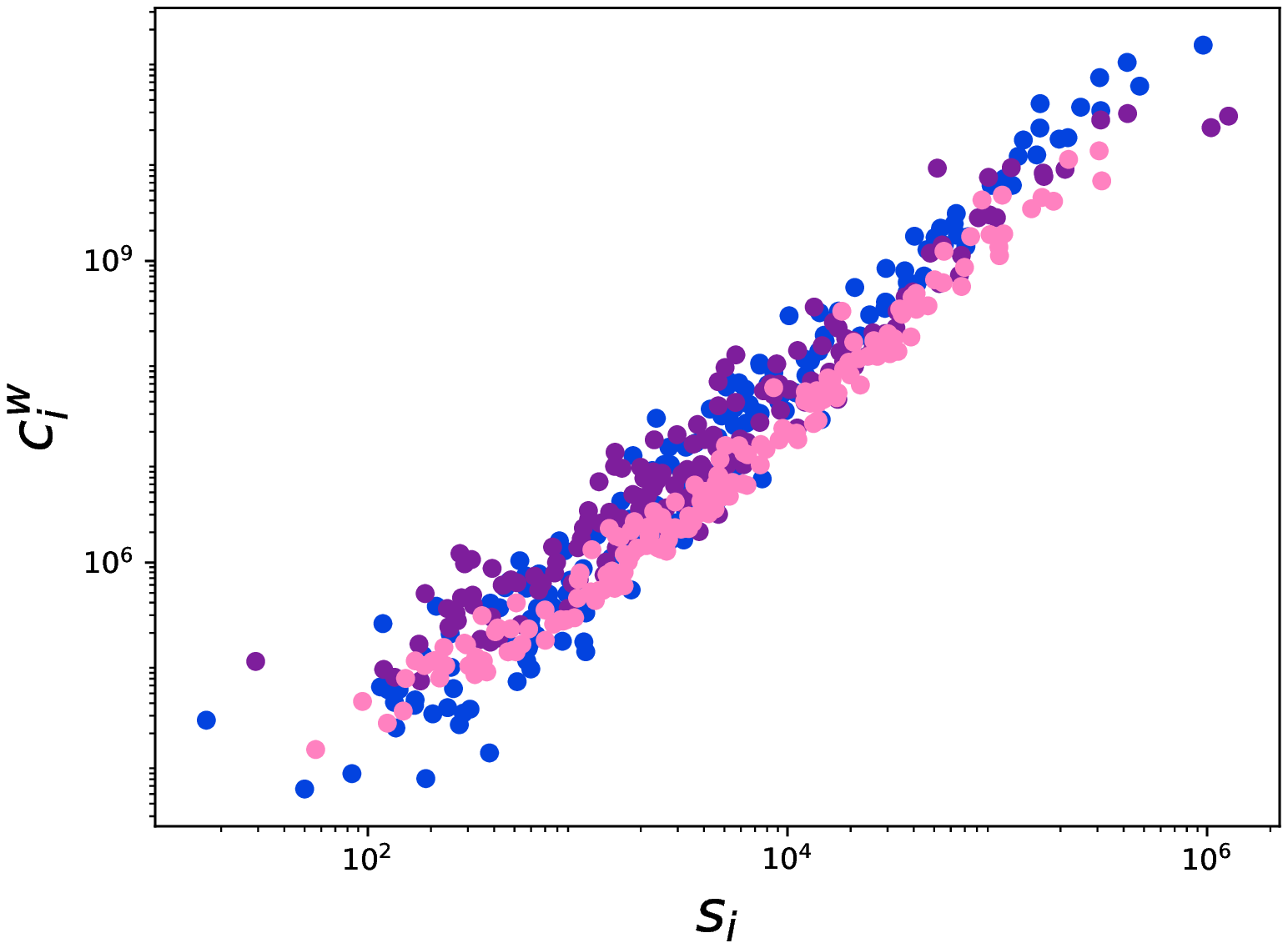}\label{fig5f}}
\caption{Performance of the negative binomial model versus the performance of the ME model described by the Hamiltonian $H_{(2)}$, in reproducing: (a) the degree distribution; (b) the ANND; (c) the BCC; (d) the strength distribution; (e) the ANNS; (f) the WCC. While a look at the Tab.~\ref{tab1} suggests the negative binomial to be the econometric model performing best, this is definitely not the case as explicitly plotting its predictions against the empirical trends reveals. Notice in (d) that the strength distributions for $H_{(2)}$ and negative binomial overlap for a large range of strength values, whereas for other statistics $H_{(2)}$ produces the best fit confirming the importance of a good topological estimation, resulting in the ability to account for the variability of observations without overfitting as can be seen in (e). Empirical points are indicated $\textcolor{xkcdBlue}{\bullet}$. Econometric and ME models are indicated as follows: $\textcolor{xkcdPink}{\bullet}$ - Negative Binomial; $\textcolor{xkcdPurple}{\bullet}$ - $H_{(2)}$. Results refer to the year 2000 of the dataset curated by Gleditsch~\cite{Gleditsch2002}.}
\label{fig5}
\end{figure*}

Some of the problems of purely econometric models are solved by looking at a different class of statistical models, i.e. the physics-inspired ones. In particular, the model described by the Hamiltonian $H_{(2)}=\sum_i\theta_ik_i+\psi_0 W+\sum_{i<j}\psi_{ij}w_{ij}$ provides a very accurate reconstruction while being favoured by information criteria. Remarkably, although it is defined by $N+1$ purely topological constraints, both AIC and BIC reveal that the latter are `irreducible', i.e. necessary to provide a satisfactory explanation of the network generating process. For the sake of comparison, Fig.~\ref{fig5} explicitly shows the performance of the models favoured by the adopted information criteria (i.e. the negative binomial model and its zero-inflated version) with that of the ME model described by $H_{(2)}$: it is evident that the ME model outperforms the purely econometric ones, still achieving a good ranking.

Looking at the class of ME models in more detail, our analysis indicates that the information carried by the strengths is not as `fundamental' as the one carried by the degrees: this is evident upon considering that 1) the UECM is always disfavoured with respect to the models just constraining the degrees, 2) the second best performing ME model is (always) the two step one, defined by a first purely topological step, accounting for the degrees, followed by an econometric-wise estimation of the weights. On top of that, we explicitly notice that structural topological information (e.g. the one provided by the link density or the degree sequence) usually correlates with node-specific economic covariates; hence, excluding such information from the model may lead to the so-called `omitted variable' bias. As proved by AIC and BIC, ME models represent the best compromise between goodness-of-fit and parsimony: in fact, they allow for structural information to be included, keeping the aforementioned type of bias low while leading to a better description of economic systems than that provided by traditional econometric models.

Our findings may indicate a route towards reconciling econometric and maximum-entropy network approaches, suggesting how to build a model that combines the pros of both: the importance of purely structural information (highlighted by physics-inspired models) can be accounted for by a model with a first step that is purely topological in nature (notice, in fact, that the TSF model is disfavoured with respect to the TS one) and a second step that takes care of estimating the weighted structure. Such an estimation can rest upon econometric considerations, driving the re-parametrization of otherwise purely structural models.

\section*{Acknowledgements}

D.G. acknowledges support from the Dutch Econophysics Foundation (Stichting Econophysics, Leiden, the Netherlands). T.S. and D.G. also acknowledge support from the European Union Horizon 2020 Program under the scheme `INFRAIA-01-2018-2019 - Integrating Activities for Advanced Communities', Grant Agreement n. 871042, `SoBigData++: European Integrated Infrastructure for Social Mining and Big Data Analytics'.

\renewcommand{\theequation}{A\arabic{equation}}

\setcounter{equation}{0}

\section*{Appendix A: Estimating the GM parameters}

Here, we consider two different specifications of the GM, i.e.

\begin{equation}
\langle w_{ij}\rangle_\text{GM}^{(1)}=\rho(\omega_i\omega_j)d_{ij}^{-1}
\end{equation}
and

\begin{equation}
\langle w_{ij}\rangle_\text{GM}^{(2)}=\rho(\omega_i\omega_j)^\beta d_{ij}^\gamma;
\end{equation}
the parameters appearing in both specifications of the GM can be estimated by implementing a Non linear-Least-Squares (NLS) regression, i.e. by solving the optimization problem

\begin{eqnarray}
&&\text{arg\:min}_{\underline{\theta}}\left\{\sum_{i<j}\left[w_{ij}-\langle w_{ij}\rangle_\text{GM}\right]^2\right\}
\end{eqnarray}
which, in turn, translates into solving the equation

\begin{eqnarray}
\sum_{i<j}\left[w_{ij}-\langle w_{ij}\rangle_\text{GM}\right]\frac{\partial\langle w_{ij}\rangle_\text{GM}}{\partial\theta_i}=0,\quad\forall\:i;
\end{eqnarray}
however, such a procedure is known to produce biased estimations. In fact, it leads to the set of conditions

\begin{equation}
\sum_{i<j}\left[w_{ij}-\langle w_{ij}\rangle_\text{GM}^{(1)}\right]\langle w_{ij}\rangle_\text{GM}=0
\end{equation}
for the first GM specification and

\begin{equation}
\begin{cases}
\sum_{i<j}\left[w_{ij}-\langle w_{ij}\rangle_\text{GM}^{(2)}\right]\langle w_{ij}\rangle_\text{GM}&=0\\
\sum_{i<j}\left[w_{ij}-\langle w_{ij}\rangle_\text{GM}^{(2)}\right]\langle w_{ij}\rangle_\text{GM}\ln(\omega_i\omega_j)&=0\\
\sum_{i<j}\left[w_{ij}-\langle w_{ij}\rangle_\text{GM}^{(2)}\right]\langle w_{ij}\rangle_\text{GM}\ln d_{ij}&=0
\end{cases}
\end{equation}
for the second GM specification. Since the conditions above are known to `weigh' more larger weights, Silva and Tenreyro~\cite{Silva2006} propose to employ a Poisson Pseudo-Maximum Likelihood (PPML) estimator, leading to the set of conditions

\begin{equation}
\sum_{i<j}\left[w_{ij}-\langle w_{ij}\rangle_\text{GM}^{(1)}\right]=0
\end{equation}
for the first GM specification and

\begin{equation}\label{eq1}
\begin{cases}
\sum_{i<j}\left[w_{ij}-\langle w_{ij}\rangle_\text{GM}^{(2)}\right]&=0\\
\sum_{i<j}\left[w_{ij}-\langle w_{ij}\rangle_\text{GM}^{(2)}\right]\ln (\omega_i\omega_j)&=0\\
\sum_{i<j}\left[w_{ij}-\langle w_{ij}\rangle_\text{GM}^{(2)}\right]\ln d_{ij}&=0.
\end{cases}
\end{equation}
for the second GM specification. Remarkably, the PPML estimator lets the purely topological condition $W=\sum_{i<j}w_{ij}=\sum_{i<j}\langle w_{ij}\rangle_\text{GM}=\langle W\rangle_\text{GM}$ (i.e. the preservation of the total weight) to be recovered for both GM specifications.

\renewcommand{\theequation}{B\arabic{equation}}

\setcounter{equation}{0}

\section*{Appendix B: Econometric models}

This Appendix is devoted to the detailed description of the econometric models considered in the present work. When coming to estimate the parameters entering into the definition of any of the econometric models considered above, we invoke the maximum-of-the-likelihood principle, prescribing to maximize the function

\begin{equation}
\mathcal{L}=\ln Q(\mathbf{W})
\end{equation}
where the optimization is carried out with respect to the set of parameters characterizing each specific model.\\

In the present contribution we have considered undirected networks, hence the parameters associated to node-specific regressors of the same economic variable (e.g. the GDPs of the country of origin and destination) are equal. In such a framework we have employed the specifications $z_{ij}=\rho(\omega_i\omega_j)^{\beta}d_{ij}^{\gamma}$ and $G_{ij}=\delta(\omega_i\omega_j)$ where $\omega_i=\frac{\text{GDP}_i}{\overline{\text{GDP}}}$, i.e. the GDP of each country is divided by the mean value of all GDPs. We would also like to stress that the parameters entering into the definition of $z_{ij}$ are equal to those employed for the standard GM specification, i.e. $z_{ij}=e^{\underline{X}\cdot\underline{\theta}}$, as evident upon taking as regressors the natural logarithm of the GDPs and that of the geographic distance, i.e.

\begin{eqnarray}
z_{ij}=e^{\underline{X}\cdot\underline{\theta}}=e^{\beta\ln\omega_i+\beta\ln\omega_j+\gamma\ln d_{ij}+c}=\rho(\omega_i\omega_j)^\beta d_{ij}^\gamma\nonumber\\
\end{eqnarray}
having posed $c=\ln\rho$.

\subsection{Poisson Model}

The probability mass function of the Poisson model reads

\begin{equation}
q_{ij}^\text{Pois}(w_{ij})=\frac{z_{ij}^{w_{ij}}e^{-z_{ij}}}{w_{ij}!};
\end{equation}
where $z_{ij}=\rho(\omega_i\omega_j)^{\beta}d_{ij}^{\gamma}$. In what follows, we will substitute $w_{ij}!$ with $\Gamma[w_{ij}+1]$, as routinely done in the packages for solving econometric models. Its log-likelihood reads

\begin{equation}
\mathcal{L}_\text{Pois}=\sum_{i<j}[w_{ij}\ln z_{ij}-z_{ij}-\ln\Gamma[w_{ij}+1]]
\end{equation}
whose optimization leads to the set of equations

\begin{equation}
\sum_{i<j}\left[\frac{w_{ij}}{z_{ij}}-1\right]\frac{\partial z_{ij}}{\partial\theta_i}=0,\quad\forall\:i
\end{equation}
reading, more explicitly,

\begin{equation}
\begin{cases}
\sum_{i<j}\left[w_{ij}-\langle w_{ij}\rangle_\text{Pois}\right]&=0\\
\sum_{i<j}\left[w_{ij}-\langle w_{ij}\rangle_\text{Pois}\right]\ln(\omega_i\omega_j)&=0\\
\sum_{i<j}\left[w_{ij}-\langle w_{ij}\rangle_\text{Pois}\right]\ln d_{ij}&=0;
\end{cases}
\end{equation}
notice that the set of equations above coincides with eqs. (\ref{eq1}). The Poisson model remains the most used one because it ensures that a network total weight is reproduced - a desirable feature to correctly estimate the weights of a network, as also observed for the `plain' GM.

\subsection{Negative Binomial Model}

The probability mass function of the Negative Binomial model reads

\begin{equation}
q_{ij}^\text{NB}(w_{ij})=\binom{m+w_{ij}-1}{w_{ij}}\left(\frac{1}{1+\alpha z_{ij}}\right)^m\left(\frac{\alpha z_{ij}}{1+\alpha z_{ij}}\right)^{w_{ij}}
\end{equation}
where $\alpha=m^{-1}$ to handle overdispersion and $z_{ij}=\rho(\omega_i\omega_j)^{\beta}d_{ij}^{\gamma}$. By replacing each binomial coefficient with the corresponding Gamma function, one recovers the expression

\begin{eqnarray}
\mathcal{L}_\text{NB}&=&\sum_{i<j}[w_{ij}\ln(\alpha z_{ij})-(m+w_{ij})\ln(1+\alpha z_{ij})\nonumber\\
&&+\ln\Gamma[m+w_{ij}]-\ln\Gamma[w_{ij}+1]-\ln\Gamma[m]];\nonumber\\
\end{eqnarray}
its optimization leads to the set of equations

\begin{equation}
\sum_{i<j}\left[\frac{w_{ij}-z_{ij}}{z_{ij}(1+\alpha z_{ij})}\right]\frac{\partial z_{ij}}{\partial\theta_i}=0,\quad\forall\:i
\end{equation}
and

\begin{equation}
\sum_{i<j}\left[\frac{w_{ij}-z_{ij}}{\alpha(1+\alpha z_{ij})}-m^2\left(\frac{\Gamma^{'}[m+w_{ij}]}{\Gamma[m+w_{ij}]}-\frac{\Gamma^{'}[m]}{\Gamma[m]}\right)\right]=0
\end{equation}
reading, more explicitly,

\begin{equation}
\begin{cases}
\sum_{i<j}\left[\frac{w_{ij}-\langle w_{ij}\rangle_\text{NB}}{\text{Var}_\text{NB}[w_{ij}]}\right]&=0\\
\sum_{i<j}\left[\frac{w_{ij}-\langle w_{ij}\rangle_\text{NB}}{\text{Var}_\text{NB}[w_{ij}]}\right]
\ln(\omega_i\omega_j)&=0\\
\sum_{i<j}\left[\frac{w_{ij}-\langle w_{ij}\rangle_\text{NB}}{\text{Var}_\text{NB}[w_{ij}]}\right]
\ln d_{ij}&=0\\
\sum_{i<j}\left[\frac{w_{ij}-\langle w_{ij}\rangle_\text{NB}}{\alpha(1+\alpha z_{ij})}-m^2\left(\frac{\Gamma^{'}[m+w_{ij}]}{\Gamma[m+w_{ij}]}-\frac{\Gamma^{'}[m]}{\Gamma[m]}\right)\right]&=0.
\end{cases}
\end{equation}

Notice that the negative binomial model does not constrain the total weight of a network, whence its bad performance in reproducing the other weighted structural properties.

\subsection{Zero-Inflated Poisson Model}

The zero-inflated version of the Poisson model is defined by a functional form reading

\begin{eqnarray}
Q(\mathbf{W})&=&\prod_{i<j}q_{ij}(w_{ij})\nonumber\\
&=&\prod_{i<j}p_{ij}^{a_{ij}}(1-p_{ij})^{1-a_{ij}}\cdot q_{ij}(w_{ij}|a_{ij})\nonumber\\
&=&P(\mathbf{A})Q(\mathbf{W}|\mathbf{A})
\end{eqnarray}
and inducing the following log-likelihood

\begin{eqnarray}
\mathcal{L}&=&\ln Q(\mathbf{W})=\ln P(\mathbf{A})+\ln Q(\mathbf{W}|\mathbf{A})\nonumber\\
&=&\sum_{i<j}[a_{ij}\ln p_{ij}+(1-a_{ij})\ln(1-p_{ij})\nonumber\\
&&+\ln q_{ij}(w_{ij}|a_{ij})]
\end{eqnarray}
where

\begin{eqnarray}
p_{ij}^\text{ZIP}&=&\frac{G_{ij}}{1+G_{ij}}(1-e^{-z_{ij}}),\\
q_{ij}^\text{ZIP}(w_{ij}|a_{ij})&=&\left[\frac{z_{ij}^{w_{ij}}e^{-z_{ij}}}{(1-e^{-z_{ij}})w_{ij}!}\right]^{a_{ij}}
\end{eqnarray}
with $z_{ij}=\rho(\omega_i\omega_j)^{\beta}d_{ij}^{\gamma}$ and $G_{ij}=\delta\omega_i\omega_j$; hence,

\begin{eqnarray}
\mathcal{L}_\text{ZIP}&=&\sum_{i<j}[a_{ij}\ln G_{ij}-a_{ij}\ln(1+G_{ij}e^{-z_{ij}})\nonumber\\
&&+\ln(1+G_{ij}e^{-z_{ij}})-\ln(1+G_{ij})\nonumber\\
&&+w_{ij}\ln z_{ij}-a_{ij}z_{ij}-a_{ij}\ln\Gamma[w_{ij}+1]].\nonumber\\
\end{eqnarray}

Its optimization leads to the set of equations

\begin{equation}
\begin{cases}
\sum_{i<j}\left[\frac{a_{ij}-p_{ij}^\text{ZIP}}{1+G_{ij}e^{-z_{ij}}}\right]
&=0\\
\sum_{i<j}
\left[w_{ij}-\left(\frac{a_{ij}+G_{ij}e^{-z_{ij}}}{1+G_{ij}e^{-z_{ij}}}\right)z_{ij}\right]
&=0\\
\sum_{i<j}
\left[w_{ij}-\left(\frac{a_{ij}+G_{ij}e^{-z_{ij}}}{1+G_{ij}e^{-z_{ij}}}\right)z_{ij}\right]\ln(\omega_i\omega_j)
&=0\\
\sum_{i<j}
\left[w_{ij}-\left(\frac{a_{ij}+G_{ij}e^{-z_{ij}}}{1+G_{ij}e^{-z_{ij}}}\right)z_{ij}\right]\ln d_{ij}
&=0
\end{cases}
\end{equation}
with a clear meaning of the symbols.

\subsection{Zero-Inflated Negative Binomial Model}

As for the ZIP model, the zero-inflated version of the negative binomial model induces a log-likelihood reading

\begin{eqnarray}
\mathcal{L}&=&\ln Q(\mathbf{W})=\ln P(\mathbf{A})+\ln Q(\mathbf{W}|\mathbf{A})\nonumber\\
&=&\sum_{i<j}[a_{ij}\ln p_{ij}+(1-a_{ij})\ln(1-p_{ij})\nonumber\\
&&+\ln q_{ij}(w_{ij}|a_{ij})]
\end{eqnarray}
where

\begin{widetext}
\begin{eqnarray}
p_{ij}^\text{ZINB}&=&\frac{G_{ij}}{1+G_{ij}}(1-\tau_{ij}),\\
q_{ij}^\text{ZINB}(w_{ij}|a_{ij})&=&\binom{m+w_{ij}-1}{w_{ij}}^{a_{ij}}\left(\frac{\tau_{ij}}{1-\tau_{ij}}\right)^{a_{ij}}\left(\frac{\alpha z_{ij}}{1+\alpha z_{ij}}\right)^{w_{ij}}\nonumber\\
&=&\left[\binom{m+w_{ij}-1}{w_{ij}}\left(\frac{1}{1-\tau_{ij}}\right)\left(\frac{1}{1+\alpha z_{ij}}\right)^m\left(\frac{\alpha z_{ij}}{1+\alpha z_{ij}}\right)^{w_{ij}}\right]^{a_{ij}}
\end{eqnarray}
\end{widetext}

with $\alpha=m^{-1}$, $\tau_{ij}=\left(\frac{1}{1+\alpha z_{ij}}\right)^m$, $z_{ij}=\rho(\omega_i\omega_j)^{\beta}d_{ij}^{\gamma}$ and $G_{ij}=\delta\omega_i\omega_j$. Hence,

\begin{eqnarray}
\mathcal{L}_\text{ZINB}&=&\sum_{i<j}[a_{ij}\ln G_{ij}-a_{ij}\ln(1+G_{ij}\tau_{ij})\nonumber\\
&&+\ln(1+G_{ij}\tau_{ij})-\ln(1+G_{ij})\nonumber\\
&&+w_{ij}\ln(\alpha z_{ij})-w_{ij}\ln(1+\alpha z_{ij})\nonumber\\
&&-ma_{ij}\ln(1+\alpha z_{ij})+a_{ij}\ln\Gamma[m+w_{ij}]\nonumber\\
&&-a_{ij}\ln\Gamma[w_{ij}+1]-a_{ij}\ln\Gamma[m]]
\end{eqnarray}
whose optimization leads to the set of equations

\begin{widetext}
\begin{equation}
\begin{cases}
\sum_{i<j}\left[\frac{a_{ij}-p_{ij}^\text{ZINB}}{1+G_{ij}\tau_{ij}}\right]&=0\\
\sum_{i<j}\left[w_{ij}-\left(\frac{a_{ij}+G_{ij}\tau_{ij}}{1+G_{ij}\tau_{ij}}\right)\right]\left(\frac{z_{ij}}{1+\alpha z_{ij}}\right)
&=0\\
\sum_{i<j}\left[w_{ij}-\left(\frac{a_{ij}+G_{ij}\tau_{ij}}{1+G_{ij}\tau_{ij}}\right)\right]\left(\frac{z_{ij}}{1+\alpha z_{ij}}\right)\ln(\omega_i\omega_j)
&=0\\
\sum_{i<j}\left[w_{ij}-\left(\frac{a_{ij}+G_{ij}\tau_{ij}}{1+G_{ij}\tau_{ij}}\right)\right]\left(\frac{z_{ij}}{1+\alpha z_{ij}}\right)\ln d_{ij}
&=0\\
\sum_{i<j}\left[\left(\frac{a_{ij}+ G_{ij}\tau_{ij}}{1+G_{ij}\tau_{ij}}\right)\left(m^2 
\ln\left(1+\alpha z_{ij}\right) - \frac{mz_{ij}}{1+\alpha z_{ij}}\right) + \dfrac{w_{ij}}{\alpha(1+\alpha z_{ij})}-m^2a_{ij}\left(\frac{\Gamma^{'}[m+w_{ij}]}{\Gamma[m+w_{ij}]}-\frac{\Gamma^{'}[m]}{\Gamma[m]}\right)
\right]&=0.
\end{cases}
\end{equation}
\end{widetext}

\renewcommand{\theequation}{C\arabic{equation}}

\setcounter{equation}{0}

\section*{Appendix C: Maximum-Entropy Models}

This Appendix is devoted to the detailed description of the maximum-entropy models considered in the present work. As for the econometric ones, the estimation of the set of parameters defining each model is carried out by maximizing the corresponding log-likelihood function, $\mathcal{L}=\ln Q(\mathbf{W})$.

ME models are derived by maximizing Shannon entropy under a suitable set of constraints. The generic probability mass function reads

\begin{eqnarray}\label{full}
Q(\mathbf{W})&=&\prod_{i<j}q_{ij}(w_{ij})\nonumber\\
&=&\prod_{i<j}p_{ij}^{a_{ij}}(1-p_{ij})^{1-a_{ij}}\cdot q_{ij}(w_{ij}|a_{ij})\nonumber\\
&=&\prod_{i<j}p_{ij}^{a_{ij}}(1-p_{ij})^{1-a_{ij}}\cdot y_{ij}^{w_{ij}-a_{ij}}(1-y_{ij})^{a_{ij}}\nonumber\\
&=&P(\mathbf{A})Q(\mathbf{W}|\mathbf{A})
\end{eqnarray}
and induces the log-likelihood

\begin{eqnarray}
\mathcal{L}&=&\ln Q(\mathbf{W})=\ln P(\mathbf{A})+\ln Q(\mathbf{W}|\mathbf{A})\nonumber\\
&=&\sum_{i<j}[a_{ij}\ln p_{ij}+(1-a_{ij})\ln(1-p_{ij})\nonumber\\
&&+\ln q_{ij}(w_{ij}|a_{ij})].
\end{eqnarray}

In order to turn maximum-entropy models into proper econometric ones, we have posed $\frac{y_{ij}}{1-y_{ij}}=z_{ij}=\rho(\omega_i\omega_j)^{\beta}d_{ij}^{\gamma}$, with $\omega_i=\frac{\text{GDP}_i}{\overline{\text{GDP}}}$.\\

Let us start instantiating the ME formalism by considering the Hamiltonian

\begin{equation}
H_{(1)}(\mathbf{W})=\theta_0L+\psi_0W+\sum_{i<j}\psi_{ij}w_{ij}
\end{equation}
that leads to

\begin{eqnarray}
\mathcal{L}_{(1)}&=&L\ln x+W\ln y_0\nonumber\\
&&+\sum_{i<j}[w_{ij}\ln z_{ij}-w_{ij}\ln(1+z_{ij})+\ln(1+z_{ij}-y_0z_{ij})\nonumber\\
&&-\ln(1+z_{ij}-y_0z_{ij}+xy_0z_{ij})]
\end{eqnarray}
whose optimization leads to solve the following set of equations

\begin{equation}
\begin{cases}
\sum_{i<j}[a_{ij}-p_{ij}^{(1)}]&=0\\
\sum_{i<j}[w_{ij}-\langle w_{ij}\rangle_{(1)}]&=0\\
\sum_{i<j}[w_{ij}-\langle w_{ij}\rangle_{(1)}]\left(\frac{1}{1+z_{ij}}\right)&=0\\
\sum_{i<j}[w_{ij}-\langle
w_{ij}\rangle_{(1)}]\left(\frac{\ln(\omega_i\omega_j)}{1+z_{ij}}\right)&=0\\
\sum_{i<j}[w_{ij}-\langle
w_{ij}\rangle_{(1)}]\left(\frac{\ln d_{ij}}{1+z_{ij}}\right)&=0\\
\end{cases}
\end{equation}
the first equation guaranteeing that the total number of links is preserved, i.e. 

\begin{equation}
L=\langle L\rangle=\sum_{i<j}\frac{xy_0z_{ij}}{1+z_{ij}-y_0z_{ij}+xy_0z_{ij}}=\sum_{i<j}p_{ij}^{(1)}
\end{equation}
and the second equation guaranteeing that the total weight is preserved, i.e. 

\begin{equation}
W=\langle W\rangle=\sum_{i<j}\frac{p_{ij}^{(1)}(1+z_{ij})}{1+z_{ij}-y_0z_{ij}}=\sum_{i<j}\langle w_{ij}\rangle_{(1)}.
\end{equation}

On the other hand, the Hamiltonian

\begin{equation}
H_{(2)}(\mathbf{W})=\sum_i\theta_ik_i+\psi_0W+\sum_{i<j}\psi_{ij}w_{ij}
\end{equation}
leads to the expression

\begin{eqnarray}
\mathcal{L}_{(2)}&=&\sum_ik_i\ln x_i+W\ln y_0\nonumber\\
&&+\sum_{i<j}[w_{ij}\ln z_{ij}-w_{ij}\ln(1+z_{ij})+\ln(1+z_{ij}-y_0z_{ij})\nonumber\\
&&-\ln(1+z_{ij}-y_0z_{ij}+x_ix_jy_0z_{ij})]
\end{eqnarray}
whose optimization requires to solve the set of equations below

\begin{equation}
\begin{cases}
\sum_{j(\neq i)}[a_{ij}-p_{ij}^{(2)}]&=0,\quad\forall\:i\\
\sum_{i<j}[w_{ij}-\langle w_{ij}\rangle_{(2)}]&=0\\
\sum_{i<j}[w_{ij}-\langle w_{ij}\rangle_{(2)}]\left(\frac{1}{1+z_{ij}}\right)&=0\\
\sum_{i<j}[w_{ij}-\langle
w_{ij}\rangle_{(2)}]\left(\frac{\ln(\omega_i\omega_j)}{1+z_{ij}}\right)&=0\\
\sum_{i<j}[w_{ij}-\langle
w_{ij}\rangle_{(2)}]\left(\frac{\ln d_{ij}}{1+z_{ij}}\right)&=0\\
\end{cases}
\end{equation}
the first set of $N$ equations guaranteeing that each degree is preserved, i.e.

\begin{equation}
k_i=\langle k_i\rangle=\sum_{j(\neq i)}\frac{x_ix_jy_0z_{ij}}{1+z_{ij}-y_0z_{ij}+x_ix_jy_0z_{ij}}=\sum_{j(\neq i)}p_{ij}^{(2)}
\end{equation}
and the $(N+1)$-th equation guaranteeing that the total weight is preserved, i.e. 

\begin{equation}
W=\langle W\rangle=\sum_{i<j}\frac{p_{ij}^{(2)}(1+z_{ij})}{1+z_{ij}-y_0z_{ij}}=\sum_{i<j}\langle w_{ij}\rangle_{(2)}.
\end{equation}

Writing a probability mass function in the factorized form $Q(\mathbf{W})=P(\mathbf{A})Q(\mathbf{W}|\mathbf{A})$ allows us to design two-step models, i.e. network models whose binary estimation step can be defined independently from the weighted one. To this aim, let us combine the probability distribution of the Undirected Binary Configuration Model, inducing a single-link probability coefficient reading $p_{ij}^\text{UBCM}=\frac{x_ix_j}{1+x_ix_j}$, with the usual weighted, conditional one, i.e. $q_{ij}(w_{ij}|a_{ij})=\left(\frac{y_0z_{ij}}{1+z_{ij}}\right)^{w_{ij}-a_{ij}}\cdot\left(\frac{1+z_{ij}-y_0z_{ij}}{1+z_{ij}}\right)^{a_{ij}}$. As a result, one obtains

\begin{eqnarray}
\mathcal{L}_\text{TS}&=&\sum_ik_i\ln x_i+(W-L)\ln y_0\nonumber\\
&&+\sum_{i<j}[-\ln(1+x_ix_j)+(w_{ij}-a_{ij})\ln z_{ij}\nonumber\\
&&-w_{ij}\ln (1+z_{ij})+a_{ij}\ln(1+z_{ij}-y_0z_{ij})]\nonumber\\
\end{eqnarray}
whose optimization requires to solve the set of equations below

\begin{equation}
\begin{cases}
\sum_{j(\neq i)}[a_{ij}-p_{ij}^\text{TS}]&=0,\quad\forall\:i\\
\sum_{i<j}[w_{ij}-\langle w_{ij}\rangle_\text{TS}]&=0\\
\sum_{i<j}[w_{ij}-\langle w_{ij}\rangle_\text{TS}]\left(\frac{1}{1+z_{ij}}\right)&=0\\
\sum_{i<j}[w_{ij}-\langle
w_{ij}\rangle_\text{TS}]\left(\frac{\ln(\omega_i\omega_j)}{1+z_{ij}}\right)&=0\\
\sum_{i<j}[w_{ij}-\langle
w_{ij}\rangle_\text{TS}]\left(\frac{\ln d_{ij}}{1+z_{ij}}\right)&=0\\
\end{cases}
\end{equation}
the first set of $N$ equations guaranteeing that each degree is preserved, i.e.

\begin{equation}
k_i=\langle k_i\rangle=\sum_{j(\neq i)}\frac{x_ix_j}{1+x_ix_j}=\sum_{j(\neq i)}p_{ij}^\text{TS}=\sum_{j(\neq i)}p_{ij}^\text{UBCM}
\end{equation}
and the $(N+1)$-th equation guaranteeing that the total conditional weight is preserved, i.e. 

\begin{equation}
W=\langle W\rangle=\sum_{i<j}\frac{a_{ij}(1+z_{ij})}{1+z_{ij}-y_0z_{ij}}=\sum_{i<j}\langle w_{ij}\rangle_\text{TS}.
\end{equation}

The amount of structural information to constrain can be further reduced by imagining a two-step model whose binary estimation step encodes a larger amount of `economic' information. To this aim, let us combine the probability distribution of the density-corrected Gravity Model, inducing a single-link probability coefficient reading $p_{ij}^\text{TSF}=\frac{G_{ij}}{1+G_{ij}}$, with the usual weighted, conditional one, i.e. $q_{ij}(w_{ij}|a_{ij}=1)=\left(\frac{y_0z_{ij}}{1+z_{ij}}\right)^{w_{ij}-a_{ij}}\cdot\left(\frac{1+z_{ij}-y_0z_{ij}}{1+z_{ij}}\right)^{a_{ij}}$. As a result, one obtains

\begin{eqnarray}
\mathcal{L}_\text{TSF}&=&\sum_{i<j}[a_{ij}\ln G_{ij}-\ln(1+G_{ij})+(W-L)\ln y_0\nonumber\\
&&+(w_{ij}-a_{ij})\ln z_{ij}-w_{ij}\ln(1+z_{ij})\nonumber\\
&&+a_{ij}\ln(1+z_{ij}-y_0z_{ij})]
\end{eqnarray}
where $G_{ij}=\delta\omega_i\omega_j$. Its optimization leads to the resolution of the following set of equations

\begin{equation}
\begin{cases}
\sum_{i<j}[a_{ij}-p_{ij}^\text{TSF}]&=0\\
\sum_{i<j}[w_{ij}-\langle w_{ij}\rangle_\text{TSF}]&=0\\
\sum_{i<j}[w_{ij}-\langle w_{ij}\rangle_\text{TSF}]\left(\frac{1}{1+z_{ij}}\right)&=0\\
\sum_{i<j}[w_{ij}-\langle
w_{ij}\rangle_\text{TSF}]\left(\frac{\ln(\omega_i\omega_j)}{1+z_{ij}}\right)&=0\\
\sum_{i<j}[w_{ij}-\langle
w_{ij}\rangle_\text{TSF}]\left(\frac{\ln d_{ij}}{1+z_{ij}}\right)&=0\\
\end{cases}
\end{equation}
the first equation guaranteeing that the total number of links is preserved, i.e.

\begin{equation}
L=\langle L\rangle=\sum_{j(\neq i)}\frac{G_{ij}}{1+G_{ij}}=\sum_{j(\neq i)}p_{ij}^\text{TSF}=\sum_{j(\neq i)}p_{ij}^\text{dcGM}
\end{equation}
and the second equation guaranteeing that the total conditional weight is preserved, i.e. 

\begin{equation}
W=\langle W\rangle=\sum_{i<j}\frac{a_{ij}(1+z_{ij})}{1+z_{ij}-y_0z_{ij}}=\sum_{i<j}\langle w_{ij}\rangle_\text{TSF}.
\end{equation}

\end{document}